\documentclass[a4paper,11pt,preprintnumbers]{article}
\pdfoutput=1 \usepackage{jcappub} 
\usepackage{booktabs}
\usepackage[T1]{fontenc} 
\usepackage{color,subfigure}
\usepackage{graphicx, multirow,soul,url,amsmath,amsfonts,amssymb,mathrsfs,amsfonts}
\usepackage[all]{hypcap}
\usepackage{url}
\usepackage{bbm}
\usepackage{cancel}
\usepackage{braket}
\usepackage{subfigure}
\usepackage[normalem]{ulem}
\usepackage{slashed}
\usepackage[dvipsnames]{xcolor}
\usepackage{multicol, blindtext}
\usepackage[section]{placeins}
\usepackage[version=3]{mhchem}
\usepackage{caption}
\usepackage{lipsum}
\usepackage{hyperref}
\usepackage{float}
\usepackage{mattens}
\usepackage{mathtools}
\usepackage{footnote}
\usepackage{amsmath,amssymb,amsthm,amsxtra,overpic,bbm,bm,epsfig}
\usepackage{color,subfigure}
\usepackage[splitrule]{footmisc}
\usepackage{bbold}
\usepackage{amsmath}
\usepackage{amsbsy}
\usepackage{lipsum}
\usepackage[export]{adjustbox}
\usepackage{units}
\newcommand{\equaref}[1]{Eq.~(\ref{#1})}

\newcommand{\equassref}[3]{Eqs.~(\ref{#1}), (\ref{#2})~and~(\ref{#3})}
\newcommand{\figref}[1]{Fig.~\ref{#1}}

\newcommand{\secref}[1]{Section~\ref{#1}}
\newcommand{\appref}[1]{Appendix~\ref{#1}}
\newcommand{\tabref}[1]{Table~\ref{#1}}

\newcommand{\bq}{\begin{eqnarray}}
\newcommand{\nq}{\end{eqnarray}}

\DeclareMathOperator{\Tr}{Tr}
\newcommand{\pushright}[1]{\ifmeasuring@#1\else\omit\hfill$\displaystyle#1$\fi\ignorespaces}
\makeatother
\newcommand{\be}{\begin{equation}}

\newcommand{\ee}{\end{equation}}
\newcommand{\bea}{\begin{eqnarray}}
\newcommand{\eea}{\end{eqnarray}}
\newcommand{\ba}{\begin{aligned}}
\newcommand{\ea}{\end{aligned}}

\title{ALP Anarchy}
\author{Francesca Chadha-Day,}
\author{James Maxwell and}
\author{Jessica Turner}
\affiliation{Institute for Particle Physics Phenomenology\\
Department of Physics, Durham University\\
Durham DH1 3LE, U.K}

\emailAdd{francesca.chadha-day@durham.ac.uk}
\emailAdd{james.l.maxwell@durham.ac.uk}
\emailAdd{jessica.turner@durham.ac.uk}

\abstract{String theory models generically predict the existence of multiple axion-like particle (ALP) fields, yet the majority of both theoretical and experimental works have assumed only one ALP. In this paper, we discuss the phenomenology of systems with multiple ALPs that can undergo oscillations akin to neutrino oscillations. Motivated by this effect, we extend the ‘anarchy’  framework, which has been used to predict neutrino oscillation parameters,  to generate the parameters of many ALP systems. We explore the phenomenology of these ALP anarchy models in some of the leading ALP search strategies, including the CERN Axion Solar Telescope, magnetic white dwarfs and the gamma-ray spectra of distant blazars. We include both the ALP-photon and the ALP-electron coupling. We find that ALP anarchy models predict drastically different results than single ALP models.
}


\begin{document}

\thispagestyle{empty}
\def\thefootnote{\fnsymbol{footnote}}
\setcounter{footnote}{1}

\setcounter{page}{0}
\maketitle
\vspace{-1cm}
\flushbottom

\def\thefootnote{\arabic{footnote}}
\setcounter{footnote}{0}

%%%%%%%%%%%%%%%%%%%%%%%%%%%%%%%%%%%%%%%%%%%%%%%%%%%%%%%%%%%%%%%%%%%%%%%%%%%%%%%
%%%%%%                          Introduction                             %%%%%%
%%%%%%%%%%%%%%%%%%%%%%%%%%%%%%%%%%%%%%%%%%%%%%%%%%%%%%%%%%%%%%%%%%%%%%%%%%%%%%%
\newpage

\section{Introduction}
Axion-like particles (ALPs) have emerged as a leading candidate for physics Beyond the Standard Model. There are three key physics motivations for the existence of ALPs. First, they behave as cold dark matter  \cite{Preskill:1982cy,Abbott:1982af,Dine:1982ah,Marsh:2015xka,Chadha-Day:2021szb}. Second,
the QCD axion offers a promising solution to the strong CP problem \cite{Peccei:1977hh,Wilczek:1977pj}. Finally, string theory suggests the existence of a large number of ALPs \cite{Voloshin:2006pz, Svrcek:2006yi,Cicoli:2012sz,Broeckel:2021dpz,Demirtas:2021gsq,Gendler:2023kjt}. Unlike the QCD axion, ALPs need not couple to gluons; therefore, their mass and Standard Model couplings are not necessarily related. ALPs are pseudo-scalar fields and are singlets under the Standard Model gauge group. This determines the form of their interactions with the Standard Model fields. For example, the Lagrangian of a single ALP $\phi$ interacting with photons and electrons is given by
\begin{equation}
\mathcal{L} \supset  -\frac{1}{2} \partial^{\mu} \phi \partial_{\mu} \phi -  \frac{1}{2} m^2 \phi^2-g^{\gamma} \phi \tilde{F}^{\mu \nu} F_{\mu \nu} + \frac{g^e}{2m_e} \bar{\psi} \gamma^{\mu} \gamma_5 \psi \partial_{\mu} \phi  \,,
\end{equation}
where $g^{\gamma}$ is the coupling of $\phi$ to the electromagnetic field strength tensor, $F^{\mu\nu}$, $g^e$ is the dimensionless ALP-electron coupling, $m_e$ is the electron mass and $\psi$ denotes the electron field.

Models containing many ALPs may differ qualitatively in their phenomenology from those containing a single axion or ALP. Recent works have explored many ALP scenarios in a range of contexts \cite{Stott:2018opm,Mehta:2021pwf,Stott:2017hvl,Reig:2021ipa,Kim:2004rp,Dimopoulos:2005ac,Agrawal:2017cmd,Kitajima:2014xla,Obata:2021nql,Higaki:2014qua,Glennon:2023jsp,Gavela:2023tzu,Chadha-Day:2021uyt,Chen:2021hfq,Cyncynates:2021xzw,Gasparotto:2023psh,Cyncynates:2022wlq,Cyncynates:2023esj}. The goal of this paper is to explore the interaction of many ALP models with the Standard Model. As pointed out in Ref.~\cite{Chadha-Day:2021uyt}, this can be understood by considering the misalignment between the ALPs' mass basis and the basis in which only one linear combination of ALPs couples to electromagnetism. As shown below, this results in oscillations between the electromagnetically coupled ALP and a number of orthogonal hidden ALP states where the physics at work is analogous to the physics of neutrino oscillations. This effect has also been studied in the context of Kaluza-Klein axions \cite{Dienes:1999gw}. This paper will expand on the results of Ref.~\cite{Chadha-Day:2021uyt} by considering several new aspects of these ALP oscillations. In particular, we will introduce ALP anarchy models, similar to the anarchy approach that has been used in neutrino physics to explain the structure of the leptonic mixing matrix. We will show that these many ALP models display dramatically different phenomenology to single ALP models with the same effective couplings. The framework developed here also applies to any system with multiple axion or ALP fields.

This paper is structured as follows. In  \secref{sec:oscillations} we will introduce many ALP models and the oscillation effect. In \secref{sec:Anarchy} we will introduce ALP anarchy models. In \secref{sec:CAST}, \secref{sec:MWD} and \secref{sec:blazar} we will calculate the phenomenology of ALP anarchy models in the Cern Axion Solar Telescope, in magnetic white dwarfs and in the gamma-ray spectra of distant blazars respectively. Finally, in \secref{sec:conclusion}, we will discuss our results further and conclude.

\section{ALP oscillations}
\label{sec:oscillations}
We will consider a model containing $N$ ALPs, each coupling to both photons and electrons:
\begin{equation}
\label{eq:massBasis}
\mathcal{L} \supset \sum_{i=1}^{N} \left( -\frac{1}{2} \partial^{\mu} \phi_{i} \partial_{\mu} \phi_{i} -  \frac{1}{2} m_i^2 \phi_i^2-g^{\gamma}_i \phi_i \tilde{F}^{\mu \nu} F_{\mu \nu} + \frac{g^e_i}{2m_e} \bar{\psi} \gamma^{\mu} \gamma_5 \psi \partial_{\mu} \phi_i  \right)\,,
\end{equation}
where $\phi_i$ is an ALP field with mass $m_i$,  $g^{\gamma}_i$ is the coupling of $\phi_i$ to the electromagnetic field strength tensor, $F^{\mu\nu}$, $g^e_i$ is the dimensionless ALP-electron coupling, $m_e$ is the electron mass and $\psi$ denotes the electron field. In this paper, we will consider only ALP interactions with photons and electrons, but similar considerations would apply to other couplings. We note that \equaref{eq:massBasis} is in the mass basis. In this work,  we will consider string-motivated ultra-light ALPs, and hence, we will assume that the QCD axion (which may also emerge from string theory) is too heavy to contribute to the scenarios considered here. Furthermore, string ALPs with masses heavier than the energy scales considered below will also not contribute. It will  be convenient to rotate to the electromagnetic basis, in which only a single ALP (the `electromagnetic ALP') couples to photons:
\begin{equation}
\label{EMbasis}
\mathcal{L} \supset - \sum_i \frac{1}{2} \partial^{\mu} \phi_{i} \partial_{\mu} \phi_{i} -  \sum_{i,j} \frac{1}{2} M^{\gamma}_{i j} \phi_i \phi_j + \sum_{i} \frac{g^e_i}{2m_e} \bar{\psi} \gamma^{\mu} \gamma_5 \psi \partial_{\mu} \phi_i   -g^{\gamma} \phi_1 \tilde{F}^{\mu \nu} F_{\mu \nu}  \,,
\end{equation}
where $g^{\gamma} = \sqrt{\sum_i {g^{\gamma}_i}^2}$ is the total effective ALP-photon coupling, we have chosen $\phi_1 = {\sum_i g^{\gamma}_i \phi_i}/{g^{\gamma}}$ as the electromagnetic ALP and $\phi_i$ and $g^e_i$ have been appropriately redefined. This basis has the advantage that only $\phi_1$ interacts directly with the photon. The other ALPs, $\phi_{\{2...N\}}$, are `hidden' with respect to the electromagnetic interaction. 

For example, any ALP production via the electromagnetic interaction {\it in cases where the mass is irrelevant} will produce the electromagnetic ALP state $\phi_1$. The production rate can, in this case, be calculated simply by considering a single ALP with coupling $g^{\gamma}$ to photons. However, this does not necessarily mean we can treat the system as though there is only one ALP with coupling to photons $g^{\gamma}$. As we assume that the ALP mass states differ and that the mass and electromagnetic basis are misaligned, $\phi_1$ may mix with the hidden states,  $\phi_{\{2...N\}}$. This is an analogous effect to neutrino oscillations resulting from misalignment between the neutrinos' interaction and mass eigenbases.

Similarly,  we can define the electronic basis in which only one ALP (the `electronic ALP') couples to electrons:
\begin{equation}
\label{electronicBasis}
\mathcal{L} \supset - \sum_i \frac{1}{2} \partial^{\mu} \phi_{i} \partial_{\mu} \phi_{i} -  \sum_{i,j} \frac{1}{2} M^e_{i j} \phi_i \phi_j - \sum_{i} g^{\gamma}_i \phi_i \tilde{F}^{\mu \nu} F_{\mu \nu} + \frac{g^e}{2m_e} \bar{\psi} \gamma^{\mu} \gamma_5 \psi \partial_{\mu} \phi_1\,,
\end{equation}
where $g^{e} = \sqrt{\sum_i {g^{e}_i}^2}$ and we have now chosen $\phi_1 = {\sum_i g^{e}_i \phi_i}/g^{e}$ as the electronic ALP and $\phi_i$ and $g^{\gamma}_i$ have been appropriately redefined. As with the electromagnetic ALP, this basis has the advantage that only $\phi_1$ interacts directly with the electron while the other ALPs $\phi_{\{2...N\}}$ are `hidden' with respect to the electron interaction. 
Note that the electronic and electromagnetic ALP states are generally neither orthogonal nor colinear. Hence, in scenarios where both ALP-photon and ALP-electron interactions are relevant, we must potentially consider three different bases - the mass, the electromagnetic, and the electronic. If other interactions between the ALPs and the Standard Model are relevant, these may introduce further relevant bases.

As shown below, ALP oscillations are phenomenologically significant for many observations, particularly when we hope to detect an ALP that has propagated a large distance. However, ALP search strategies that rely only to the disappearance of Standard Model particles or energy into ALP degrees of freedom, such as stellar cooling bounds on ALPs \cite{Ayala:2014pea,Raffelt:1985nj,Blinnikov:1994eoa}, are not significantly effected by ALP oscillations. Therefore, comparison between ALP searches is substantially more complicated in many ALP systems than if we assume only a single axion or ALP.

\section{Anarchy models}\label{sec:Anarchy}

String theory provides a framework to understand ALP properties, including their mixing matrices that parameterise the misalignment between the interaction and mass bases. Calculating these mixing matrices is a highly non-trivial task \cite{Halverson:2018cio}. Nonetheless, the ALP photon coupling has been modelled in a range of string axiverse scenarios \cite{Halverson:2019cmy,Gendler:2023kjt}. Such modelling of the electronic ALP properties from string theory has not been undertaken.

This current lack of knowledge of the mixing matrices, from a first principle string theory calculation, presents a challenge when motivating the choice of ALP mixing parameters and couplings. In this work, we circumvent this issue and remain agnostic to the ALPs' particular ultraviolet physics by considering a large set of randomly sampled mixing matrices. This framework, also known as \emph{anarchy},  has been applied in neutrino physics and refers to the postulate that the neutrino mass matrix has no particular structure but that its elements are randomly chosen $\mathcal{O}(1)$ parameters  \cite{Hall:1999sn,Haba:2000be,deGouvea:2003xe,deGouvea:2012ac,Espinosa:2003qz,Heeck:2012fw,Bai:2012zn,Lu:2014cla,Fortin:2016zyf,Fortin:2017iiw,Fortin:2018etr,Fortin:2020oud}. Randomness in $\mathcal{O}(1)$ coupling constants is expected in sufficiently complicated models or with many fields mixing with each other. 
While it remains unclear if string theory predicts an anarchy-like mixing pattern, in this work, we use anarchy to explore the general properties of multi-ALP phenomenology and their relation to the number of ALP mass eigenstates.

To implement this anarchical approach and determine the mixing matrices between the hidden and visible ALPs,  we assume that the non-diagonal ALP mass matrices, $M^\gamma$ and $M^e$, are real and can, therefore, be related to the mass basis states as follows:
\begin{equation}
      M^{\alpha} = U^\alpha D  {U^\alpha}^T\,\quad \alpha = e, \gamma\,,
\end{equation}
where we assume $U^\alpha \in SO(N)$  and $D=\operatorname{diag}\left(m_1, m_2, \ldots, m_N\right)$ is a real diagonal  matrix with $m_i$ denoting the mass of ALP field $\phi_i$. Following our assumption that the electromagnetic and electronic bases are misaligned, a given model will be described by a set of two rotations $U^\alpha$ - one for each interaction basis. We sample $SO(N)$ such that elements are uniformly distributed over the group manifold to generate these mixing matrices. This can be achieved using the Haar measure, which describes the density of elements in a Lie group. In spherical coordinates, the Haar measure for $SO(N)$ is:
\begin{equation}\label{eq:Haar}
    dV = \left( \prod_{i \in [1, N-1], j \in [i, N-1]} \sin^{i-1} \theta_{i,j} \right) d\theta_{1,1} \cdots d\theta_{N-1, N-1}
       \quad
       \,,
\end{equation}
where there are $N(N-1)/2$ mixing angles, $\left\{\theta_{i j}\right\}_{1 \leq i \leq j \leq N-1}$ with $\theta_{1, j} \in [0, 2\pi]$ and $\theta_{i, j} \in [0, \pi]$ for $i, j > 1$. Sampling uniformly in $dV$ yields the desired distribution of the angles that parameterise the mixing matrices $U^{\gamma}$ and $U^{e}$ and we outline our numerical procedure for this task in  \appref{sec:AppA}. Mixing matrices, although providing a simple means to understand a given parametrisation, contain a large degree of redundancy. In practice, and in what follows, we are more interested in the relationship between the couplings in the mass basis -- $\{ g_i^\gamma \}$ and $\{ g_i^e \}$. Given, for example, the EM coupling in the EM-ALP basis, the mass basis couplings are given by:

\begin{equation}
    \begin{pmatrix}
        g_1^\gamma \\
        g_2^\gamma \\
        \vdots
    \end{pmatrix} = U^\gamma 
    \begin{pmatrix}
        g^\gamma \\
        0 \\
        \vdots
    \end{pmatrix}.
    \label{eqn:genCoups}
\end{equation}
An analogous statement is also true for the electron-ALP couplings. From \equaref{eqn:genCoups}, it is clear that the top row of $U^\gamma$ parameterises the mixing of the electromagnetic ALP with the hidden states, which will have observable phenomenological consequences. The remaining $N-1$ rows determine the mixing \emph{between} the hidden states, which is not observationally relevant.

We note that the mixing matrices $U$ parameterise only the misalignment between the electromagnetic, electronic and mass bases and {\it not} the magnitude of the total effective couplings $g^\gamma$ and $g^e$. This allows us to distinguish between the effects of basis misalignment and the effects of simply varying the total coupling, which is also present in the single ALP case. In the following sections, we will explore the phenomenology of ALP anarchy models. In particular, we will compare anarchy models with different numbers of ALPs to single ALP models with the same total effective couplings. We note here that, in a string axiverse scenario, the fundamental parameters are the couplings $g_i^{\gamma}$ and $g_i^{e}$ of the mass eigenstates and the effective couplings of the electromagnetic and electronic ALP states emerge from these. In this case, the possible values of the effective couplings $g^{\gamma}$ and $g^e$ are determined by the string model.
%%%%%%%%%%%%%%%%%%%%%%%%%%%%%%%%%%%%%%%%%%%%%%%%%%%%%%%%%%%%%%%
\section{The Cern Axion Solar Telescope}
\label{sec:CAST}
\begin {figure}[t]
\centering
\includegraphics[width=0.8\textwidth]{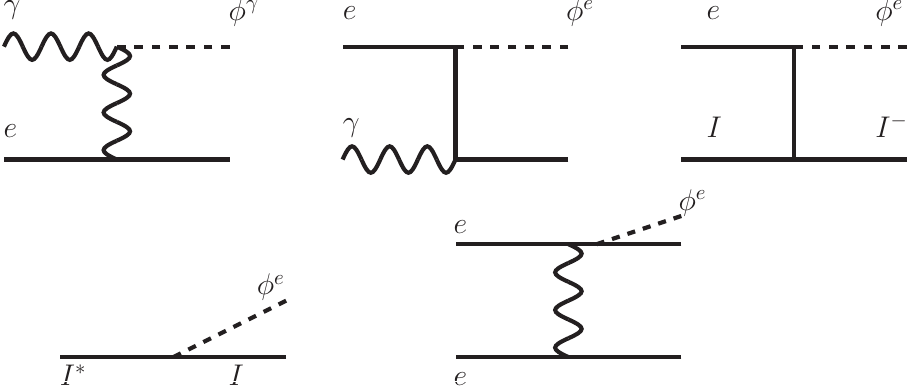}
\caption{Top left image shows the Primakoff process that produces the electromagnetic ALP, and the middle and top right images show the Compton scattering and axio-recombination processes that produce the electronic ALP. The bottom row shows axio-deexcitation and Bremsstrahlung processes that produce the electronic ALP.}\label{fig:axionprodsun}
\end{figure}

In this section, we will outline the production of electromagnetic and electronic solar ALPs in the Sun and their detection by CAST. We will discuss how oscillations, within the framework of an anarchical mixing pattern, influence the number of electromagnetic solar ALP states that reach the CAST detector. Finally, we will explain how we reinterpret the CAST experimental constraint on the single ALP parameter space for our multi-ALP scenario.
Of the potential astrophysical sources, the Sun is one the most accessible to the search for ALPs as the internal dynamics of the Sun are well understood and can be accurately modelled by a weakly coupled plasma. Within this setting, it is possible to find analytical forms for the emitted ALP flux, and these can be used to set competitive bounds on ALP couplings for mass ranges relevant to solar processes. ALPs are produced predominantly through three mechanisms: axio-bremsstrahlung (the Primakoff process) \cite{Raffelt:1985nk,Krauss:1984gm}, axio-recombination and axio-deexcitation \cite{Dimopoulos:1985tm,Dimopoulos:1986kc,Pospelov:2008jk}, and Compton scattering \cite{Mikaelian:1978jg,Fukugita:1982ep,Fukugita:1982gn}, see Ref.~\cite{Redondo:2013wwa} for a comprehensive overview. Of these processes, shown in \figref{fig:axionprodsun}, only the Primakoff process depends on the ALP-photon coupling with the others arising from the ALP-electron interaction. The CERN Axion Solar Telescope (CAST) is a helioscope designed to use the Primakoff effect to scatter axions or ALPs produced by the Sun into photons. CAST consisted of a $9.2\,\rm{m}$ long evacuated cylinder with a sustained magnetic field oriented transverse to the ALP propagation direction. A bound on $g^{\gamma}$ and $g^{e}$ can be placed based on the lack of an ALP signal. In this work, we consider how the multi-ALP scenario will alter this bound. This modification occurs as the electromagnetic and electronic ALP states produced in the Sun may oscillate into hidden ALP states which CAST cannot detect.

\subsection{Solar ALP emission and detection}
If ALPs couple to the electromagnetic field strength tensor or an electronic current, as in \equaref{eq:massBasis}, they can be produced in the Sun. The relevant ALP production processes are shown in \figref{fig:axionprodsun}, and the dominant ALP production mechanisms are Bremmstrahlung ($B$), Compton Scattering ($C$) and the Primakoff process ($P$). In the well-studied single ALP case, with no oscillation into hidden ALPs, the fluxes at Earth, in units of   $\mathrm{m}^{-2}\, \text{year}^{-1}\, \mathrm{keV}^{-1}$, generated by these mechanisms are given by \cite{Irastorza:2011gs}:
\begin{align}
   \label{eq:diffintB} \left.\frac{\mathrm{d} \Phi_a}{\mathrm{~d} \omega}\right|_B&=8.3 \times 10^{20}\left(\frac{g_{ e}}{10^{-13}}\right)^2 \frac{\omega}{1+0.667 \omega^{1.278}} e^{-0.77\omega} \,, \\
  \label{eq:diffintC}  \left.\frac{\mathrm{d} \Phi_a}{\mathrm{~d} \omega}\right|_C&=4.2 \times 10^{18}\left(\frac{g_{ e}}{10^{-13}}\right)^2 \omega^{2.987} e^{-0.776\omega} \,, \\
  \label{eq:diffintP}  \left.\frac{\mathrm{d} \Phi_a}{\mathrm{~d} \omega}\right|_P&=2.0 \times 10^{18}\left(\frac{g_{ \gamma}}{10^{-12} \mathrm{GeV}^{-1}}\right)^2 \omega^{2.450} e^{-0.829 \omega} 
\,,
\end{align}
where the ALP energy, $\omega$, is in $\mathrm{keV}$. A more detailed study of the solar ALP flux can be found in \cite{Hoof:2021mld}. 
Integrating over the energy range ($0.8 - 6.8$ keV) of the CAST analysis \cite{Barth:2013sma}, the  total fluxes  can be found for $\Phi_B$, $\Phi_C$ and $\Phi_P$:
% \be
\begin{align}
&\label{eq:intB}\frac{\Phi_{B}}{\mathrm{m}^{-2}\, \text{year}^{-1}} = 4.1 \times 10^{46} {g^e}^2\,,\\ 
&\label{eq:intC}\frac{\Phi_{C}}{\mathrm{m}^{-2}\, \text{year}^{-1}} = 5.2 \times 10^{45} {g^e}^2\,,\\ 
&\label{eq:intP}\frac{\Phi_{P}}{\mathrm{m}^{-2}\, \text{year}^{-1}} = 1.0 \times 10^{43} \left({\frac{g^\gamma}{\mathrm{GeV}^{-1}}}\right)^2\,.
\end{align}
% \ee
The CAST experimental constraints assume a single ALP that couples to electromagnetism and electrons. We will now consider the CAST signal from models with multiple ALP mass eigenstates.

\subsection{Oscillation: The two ALP case}\label{sec:2ALP}
We will expand upon the results of ~\cite{Chadha-Day:2021uyt} by discussing the basics of ALP oscillations relevant to CAST
in a simplified two ALPs scenario where $\phi_e$ and $\phi_\gamma$ are linear combinations of only two massive ALP states, $\phi_1$ and $\phi_2$, which have masses $m_1$ and $m_2$ respectively. 
 The Lagrangian in the mass basis is
\be\ba
\mathcal{L} \supset & -\frac{1}{2} \partial^\mu \phi_1 \partial_\mu \phi_1-\frac{1}{2} \partial^\mu \phi_2 \partial_\mu \phi_2  -\frac{1}{2} m_1^2 \phi_1^2-\frac{1}{2} m_2^2 \phi_2^2\\
&-g_1^\gamma \phi_1 \tilde{F}^{\mu \nu} F_{\mu \nu}-g_2^\gamma \phi_2 \tilde{F}^{\mu \nu} F_{\mu \nu} + \frac{g^e_1}{2m_e} \bar{\psi} \gamma^{\mu} \gamma_5 \psi \partial_{\mu} \phi_1+ \frac{g^e_2}{2m_e} \bar{\psi} \gamma^{\mu} \gamma_5 \psi \partial_{\mu} \phi_2\, .
\ea\ee
We can rotate to a basis where a single ALP field, the electromagnetic ALP, couples to electromagnetism while the other hidden electromagnetic ALP field does not. These are given by
\begin{equation}
\label{eq:eigenstates}
\phi_\gamma=\frac{g_1^\gamma \phi_1+g_2^\gamma \phi_2}{\sqrt{g_1^{\gamma 2}+g_2^{\gamma 2}}}\,,
\quad 
\phi_{\gamma_h}=\frac{g_2^\gamma \phi_1-g_1^\gamma \phi_2}{\sqrt{g_1^{\gamma 2}+g_2^{\gamma 2}}} \,,
\end{equation}
respectively.
Analogously,  the electronic ALP and the  hidden electronic ALP are the following orthogonal combinations:
\begin{equation}
\phi_e=\frac{g_1^e \phi_1+g_2^e \phi_2}{\sqrt{g_1^{e 2}+g_2^{e 2}}}\,,
\quad
\phi_{e_h}=\frac{g_2^e \phi_1-g_1^e \phi_2}{\sqrt{g_1^{e 2}+g_2^{e 2}}}\,,
\end{equation}
\noindent
respectively. Note that the electromagnetic and electron ALPs $\phi_\gamma$ and $\phi_e$ are generally neither colinear nor orthogonal. Therefore,  $\phi_{\gamma_h}$ {\it will} in general have some non-zero coupling to electrons and $\phi_{e_h}$ {\it will} in general have some non-zero coupling to photons. The following unitary rotation matrices relate the mass basis and the electromagnetic and electron bases:

\begin{equation}
\begin{aligned}
&\left(\begin{array}{l}
\phi_\gamma \\
\phi_{h_\gamma}
\end{array}\right)=\left(\begin{array}{cc}
\cos(\theta^\gamma) & \sin(\theta^\gamma) \\
-\sin(\theta^\gamma) & \cos(\theta^\gamma)
\end{array}\right)
\left(\begin{array}{l}
\phi_1 \\
\phi_2
\end{array}\right)\,, \\
&\left(\begin{array}{l}
\phi_e \\
\phi_{h_e}
\end{array}\right)=\left(\begin{array}{cc}
\cos(\theta^e) & \sin(\theta^e) \\
-\sin(\theta^e) & \cos(\theta^e)
\end{array}\right)
\left(\begin{array}{l}
\phi_1 \\
\phi_2
\end{array}\right)\,,
\end{aligned}
\end{equation}

\noindent
with
\begin{equation}
\cos\left(\theta^\alpha\right)=\left(\frac{g_1^\alpha}{\sqrt{g_1^{\alpha 2}+g_2^{\alpha 2}}}\right)\,,
\quad\sin\left(\theta^\alpha\right)=\left(\frac{g_2^\alpha}{\sqrt{g_1^{\alpha 2}+g_2^{\alpha 2}}}\right)\,,
\end{equation}

\noindent
where $\alpha=\gamma, e$.

We will assume that all couplings are real, and their values are determined using the anarchical approach outlined in \secref{sec:Anarchy}. We will also assume that $m_1$ and $m_2$ are much less than other relevant energy scales. In particular, we will assume $m_1, m_2 < 10^{-2}\,\rm{eV}$, corresponding to CAST bounds for evacuated magnet bores. In this mass range, the ALP masses are also much lower than their production energy in the Sun and can be treated in the relativistic limit. This means their effects on ALP production may be neglected, so axio-recombination, axio-de-excitation and Compton scattering produce the state $\phi_e$ while Primakoff production produces the state $\phi_\gamma$. Furthermore, in this mass range, the CAST sensitivity is independent of mass, and therefore, CAST will detect the state $\phi_\gamma$ as a single signal.

As CAST aims to detect electromagnetic ALPs, $\phi_\gamma$, we are interested in the probability 
that a solar electronic or electromagnetic ALP oscillates to an electromagnetic ALP after travelling a distance $L$. We can calculate these probabilities using the fact that the mass eigenstates propagate as $\ket{\phi_i(L) } = {\rm e}^{-i \frac{m_i^2 L}{2 \omega}} \ket{\phi_i(0)}$, where $L$ is the distance travelled. The former is given by
\begin{equation}\label{eq:Prob}
\ba
P(\phi_e\to\phi_\gamma)\equiv P_{e \to \gamma} &=\frac{1}{2}\left(1+\cos(2\theta^e)\cos(2\theta^\gamma)+ \sin(2\theta^e)\sin(2\theta^\gamma)\cos\left(\frac{\Delta m^2 L}{2\omega}\right)\right)\,,
\ea
\end{equation}
where  $\Delta m^2=m^2_2-m^2_1$ is the mass squared splitting between the ALP mass states. Rewriting \equaref{eq:Prob}   in terms of  couplings, $g^\alpha$, yields
\begin{equation}\label{eq:Prob2}
P_{e \to \gamma}=\frac{\left({g_1^e}^2 g_1^{\gamma 2}+g_2^{e 2} g_2^{\gamma 2}\right)}{\left({g_1^e}^2+g_2^{e 2}\right)\left(g_1^{\gamma 2}+g_2^{\gamma 2}\right)}\left(1+\frac{2 g_1^e g_1^\gamma g_2^e  g_2^\gamma}{\left({g_1^e}^2 g_1^{\gamma 2}+g_2^{e 2} g_2^{\gamma 2}\right)}\cos\left(\frac{\Delta m^2 L}{2\omega}\right)\right)\,.
\end{equation}
In addition to the electronic ALPs produced in the Sun oscillating to electromagnetic ALPs when they reach CAST, we must consider the survival probability of the solar electromagnetic ALP
which is the usual two-state survival probability familiar from neutrino physics:
\begin{equation}
\ba
P(\phi_\gamma\to\phi_\gamma)\equiv P_{\gamma \to \gamma}&=1-\sin^2(2\theta^\gamma) \sin^2\left(\frac{\Delta m^2 L}{4\omega}\right)\\
& =1-4 \frac{g_1^{\gamma 2}g_2^{\gamma 2}}{\left(g_1^{\gamma 2}+g_2^{\gamma 2}\right)^2}  \sin^2\left(\frac{\Delta m^2 L}{4\omega}\right)\,.
\ea
\end{equation}

Several simplifications can be made for solar ALP oscillations: first, the matter potential induced by the solar electron background is negligibly small and, therefore, does not affect the electron ALP propagation through the Sun. This can be estimated from the fact that the potential experienced by electron neutrinos from electrons in the Sun is $V\approx G_F N_e\sim 10^{-12}\, \mathrm{eV}$ where $G_F$ is Fermi's constant and $N_e$ is the number density of electrons in the Sun's core. In contrast, the potential experienced by the electronic ALP (with coupling $\frac{g^e}{2 m_e}=10^{-11}\,\mathrm{GeV}^{-1}$) is $V\approx \left( \frac{g^e}{2 m_e} \right)^2 N_e\sim 10^{-29} \rm{eV}$. Likewise, the potential induced by the Sun's magnetic field is negligible, so vacuum oscillation between the ALP states will be applied.
Second, for Sun-Earth distances,  keV solar ALP energies with $\Delta m^2 > 10^{-12}\,\mathrm{eV}^2$, which we assume, the oscillation probability of \equaref{eq:Prob} and \equaref{eq:Prob2} averages when integrated over the CAST energy range yielding:
\begin{equation}\label{eq:Prob3}
P_{e \to \gamma}=\frac{\left({g_1^e}^2 g_1^{\gamma 2}+g_2^{e 2} g_2^{\gamma 2}\right)}{\left({g_1^e}^2+g_2^{e 2}\right)\left(g_1^{\gamma 2}+g_2^{\gamma 2}\right)}\,,\quad
P_{\gamma \to \gamma} = \frac{{g^{\gamma}_1}^4 + {g^{\gamma}_2}^4}{\left(g_1^{\gamma 2}+g_2^{\gamma 2}\right)^2}\,.
\ee
%%%%%%%%%%%%%%%%%%%%%%%%%%%%%%%%%%%%%%%%%%%%%%%%%%%%%%%%%%%
\subsection{Oscillation: The many ALP case}\label{sec:manyALp}
We now turn to the case where many ALP mass eigenstates couple to electrons and photons:
\begin{equation} 
\mathcal{L} \supset \sum_i^N \left(\frac{1}{2} \partial^{\mu} \phi_i \partial_{\mu} \phi_i - \frac{1}{2} m_i^2 \phi_{i^2}  - g_i^{\gamma} \phi_i \tilde{F}^{\mu \nu} F_{\mu \nu} + g_i^e \bar{\psi} \gamma^{\mu} \gamma_5 \psi \partial_{\mu} \phi_i \right)\,,
\label{eqn:mulAxLag}
\end{equation}
such that the electromagnetic and electronic ALPs produced in the Sun are linear combinations of the mass states:
\begin{equation} 
\phi_{\gamma} = \frac{\sum_i g_i^{\gamma} \phi_i}{\sqrt{\sum_i {g_i^{\gamma}}^2}}\,,
\quad 
\phi_{e} = \frac{\sum_i g_i^{e} \phi_i}{\sqrt{\sum_i {g_i^{e}}^2}}\,,
\end{equation}
where we again assume that all ALP masses considered are $ m_{i}< 10^{-2}\,\rm{eV}$. Any mass eigenstates with $m_i > 10^{-2}\,\rm{eV}$ would not contribute to the signal considered here, as they would not produce a signal in CAST with an evacuated bore. Again, we will assume that $\Delta m^2 > 10^{-12}\,\mathrm{eV}^2$ so that the oscillation probabilities average when integrated over the CAST energy range.
Under these conditions, it can be shown that the electromagnetic ALP survival probability is
\begin{equation}
P_{\gamma \to \gamma} = \frac{1}{N} \left(1 + \frac{N^2 \mathrm{VAR} \left( \{ {g_i^{\gamma}}^2 \} \right)}{{g^{\gamma}}^4} \right)= \frac{\sum_{i}^{N}{g^\gamma_{i}}^4}{\left( \sum_{i}^N{g^\gamma_{i}}^2 \right)^2}\,,
\label{eqn:gg}
\end{equation}
where $N$ is the number of ALP mass eigenstates, $g^{\gamma} = \sqrt{\sum {g_i^{\gamma}}^2}$ is the coupling of the electromagnetic ALP to photons, and $\mathrm{VAR}(\{ {g^{\gamma}_i}^2 \})$ is the variance of the ALP-photon couplings squared in the mass basis which is given by
\begin{equation} 
\mathrm{VAR} \left( \{ {g_i^{\gamma}}^2 \} \right) = \frac{\sum_i^N \left[ {{g^{\gamma}_i}}^2 - \frac{{{g^{\gamma}}}^2}{N} \right]^2}{N}\,.
\end{equation}
Likewise, the probability that an electronic ALP oscillates to an electromagnetic ALP is
\begin{equation}
P_{e \to \gamma} = \frac{1}{N} \left(1 + \frac{N^2 \mathrm{COVAR} \left( \{ {g_i^{\gamma}}^2 \}, \{ {g_i^{e}}^2 \}\} \right)}{{g^{\gamma}}^2 {g^e}^2} \right) = \frac{\sum_{i}^{N}{g^e_{i}}^2{g^\gamma_{i}}^2}{\sum_{i}^{N}{g^e_{i}}^2 \sum_{i}^N{g^\gamma_{i}}^2}\,,
\label{eqn:eGammaCov}
\end{equation}
where $g^{e} = \sqrt{\sum {g_i^{e}}^2}$ is the coupling of the electronic ALP to electrons and $\mathrm{COVAR} \left( \{ {g_i^{\gamma}}^2 \}, \{ {g_i^{e}}^2 \}\} \right)$ is the covariance of the ALP-photon and ALP-electron couplings squared and is
\begin{equation} 
\mathrm{COVAR} \left( \{ {g_i^{\gamma}}^2 \}, \{ {g_i^{e}}^2 \}\} \right) = \frac{\sum_i^N \left[ {g^{\gamma}_i}^2 - \frac{{g^{\gamma}}^2}{N} \right] \left[ {g^{e}_i}^2 - \frac{{g^{e}}^2}{N} \right]}{N} \,.
\end{equation}
%%%%%%%%%%%%%%%%%%%%%%%%%%%%%
\subsection{Reintepretation of CAST results}\label{sec:CASTRes}
%%%%%%%%%%%%%%%%%%%%%%%%%%%%%
The non-observation of an excess number of photons allows CAST to place constraints on the $(g^\gamma, g^e)$ parameter space, which bounds the solar ALP fluxes on Earth. In the multi-ALP case, the fluxes of the electromagnetic and electronic ALP on Earth, namely the oscillated fluxes, are, respectively:
\be
\ba\label{eq:oscflux}
    {\Phi^{\rm{osc}}_{\gamma}} &= P_{\gamma \to \gamma} \Phi_\gamma + P_{e \to \gamma} \Phi_e \,, \\
    \Phi_{e}^{\rm{osc}} &= P_{\gamma \rightarrow e} \Phi_\gamma + P_{e \rightarrow e} \Phi_e\,,
\ea\ee
where  $\Phi_\gamma = \Phi_P$ and $\Phi_e = \Phi_B +\Phi_C$ and $P_{a \rightarrow b}$ is the probability of species $a$ oscillating into species $b$ during propagation as derived in \secref{sec:manyALp}. 
Since CAST makes use of a strong magnetic field to convert ALPs, the relevant flux to consider for the recasting from the single to the multi-ALP scenario is $\Phi_{\gamma}^{\rm{osc}}$.
CAST has placed bounds on $(g^\gamma)^2$ as a function of ALP mass and on $g^\gamma$ as a function of $g^e$, for small ALP mass $m_a$, for a single ALP. Here we determine an upper bound on $g^{\gamma}$ as a function of $g^{e}$ for multi-ALP scenarios with $m_a \leq \, 10^{-2} \, {\rm eV}$ and $\Delta m^2 > 10^{-12} \, {\rm eV}^2$ for all relevant ALP mass eigenstates;  for which the bound becomes mass independent \cite{Barth:2013sma}. We do this by considering the maximum flux compatible with the non-detection of ALPs by CAST. Given the original single ALP bound on the ALP-electromagnetic coupling, which we denote as $g^{ \gamma}_{N=1}(g^{e})$,  (shown by the black line in \figref{fig:reCAST}), the maximum flux of electromagnetic ALP on Earth is bounded to be:
\begin{equation}
     \Phi^\mathrm{Max}(g^{e}, g^{\gamma}) = \frac{(g^{\gamma}_{N=1}(g^{e}))^2}{{g^{\gamma}}^2} \Phi^{N=1}(g^{e}, g^{\gamma}_{N=1}(g^{e}))\,,
\end{equation}
\noindent
where $\Phi^{N=1} = \Phi_P + \Phi_B + \Phi_C$ is the flux at the detector in the single ALP case. With ${\Phi_{\gamma}^{\rm{osc}}}(g^{e}, g^{\gamma})$ in hand, we now perform a grid scan over the coupling parameter space, ($g^{e}, g^{\gamma}$), to determine the new bound. For a given point in the ALP anarchy parameter space to be allowed by existing CAST data, the total EM-ALP flux at the detector given in \equaref{eq:oscflux} must be less than $\Phi^\mathrm{Max}$.

To determine the bound on the total effective couplings in the ALP anarchy scenario, it is therefore necessary to compute $P_{\gamma \to \gamma}$ and $P_{e \to \gamma}$. From \secref{sec:manyALp}, these are given by \equaref{eqn:gg} and \equaref{eqn:eGammaCov}, respectively, and can be written in terms of the relationship between the individual ALP couplings ($g_i^e$ and $g_i^\gamma$). It is this relationship that encodes the mixing between the various ALP states. 
%As stated previously,  the precise form of this mixing cannot be readily extracted from string theories at present. Consequently, we rely on the anarchical approach of \secref{sec:Anarchy} for the generation of $g_i^e$ and $g_i^\gamma$ and the subsequent calculation of $P_{\gamma \to \gamma}$ and $P_{e \to \gamma}$. 
We consider $10^4$ realisations of $\{g^\gamma_i\}$ and $\{g^e_i\}$ for each overall coupling pair ($g^\gamma, g^e$). For each point in ($g^\gamma, g^e$) space, we determine the proportion of viable realisations such that $\Phi_{\gamma}^{\rm{osc}}<\Phi^\mathrm{Max}(g^{e}, g^{\gamma})$. These results are shown in \figref{fig:reCAST}; for a detailed outline of this numerical procedure, see \appref{sec:CASTBounds}. The black line indicates the original $N=1$ bound ($\phi_e\equiv \phi_\gamma$) where in the mass independent region $\log_{10}\left(g^\gamma\,[\rm{GeV}^{-1}]\right)\lesssim -10$.
The boundary between regions where $0$ and $100 \%$ of realisations are viable can be interpreted as the bound on $g^\gamma$ as a function of $g^e$ in the ALP anarchy scenario.
%%%%%%%%%%%%%%
\begin{figure}
  \centering
  \includegraphics[width=0.45\textwidth]{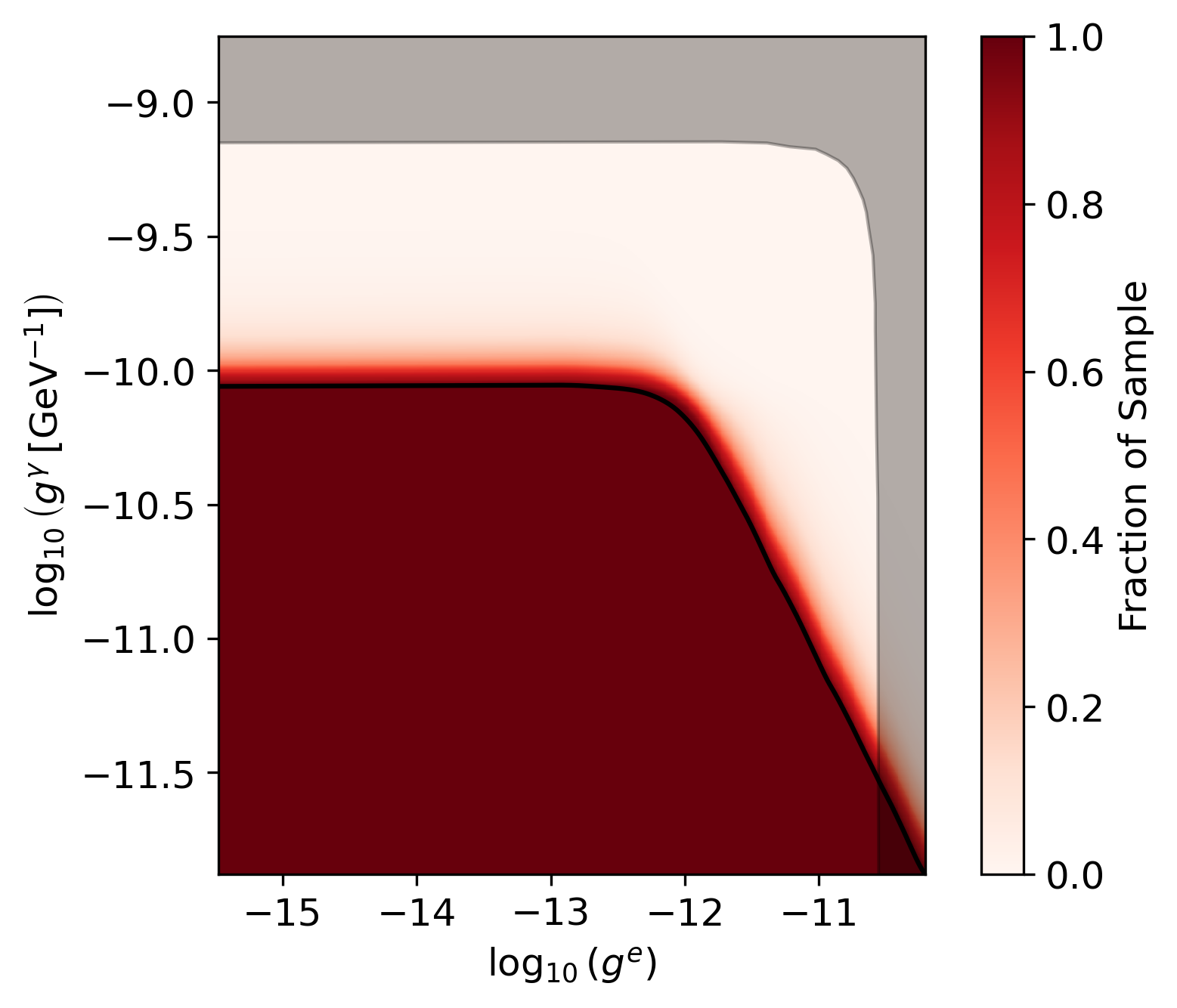}
  \includegraphics[width=0.45\textwidth]{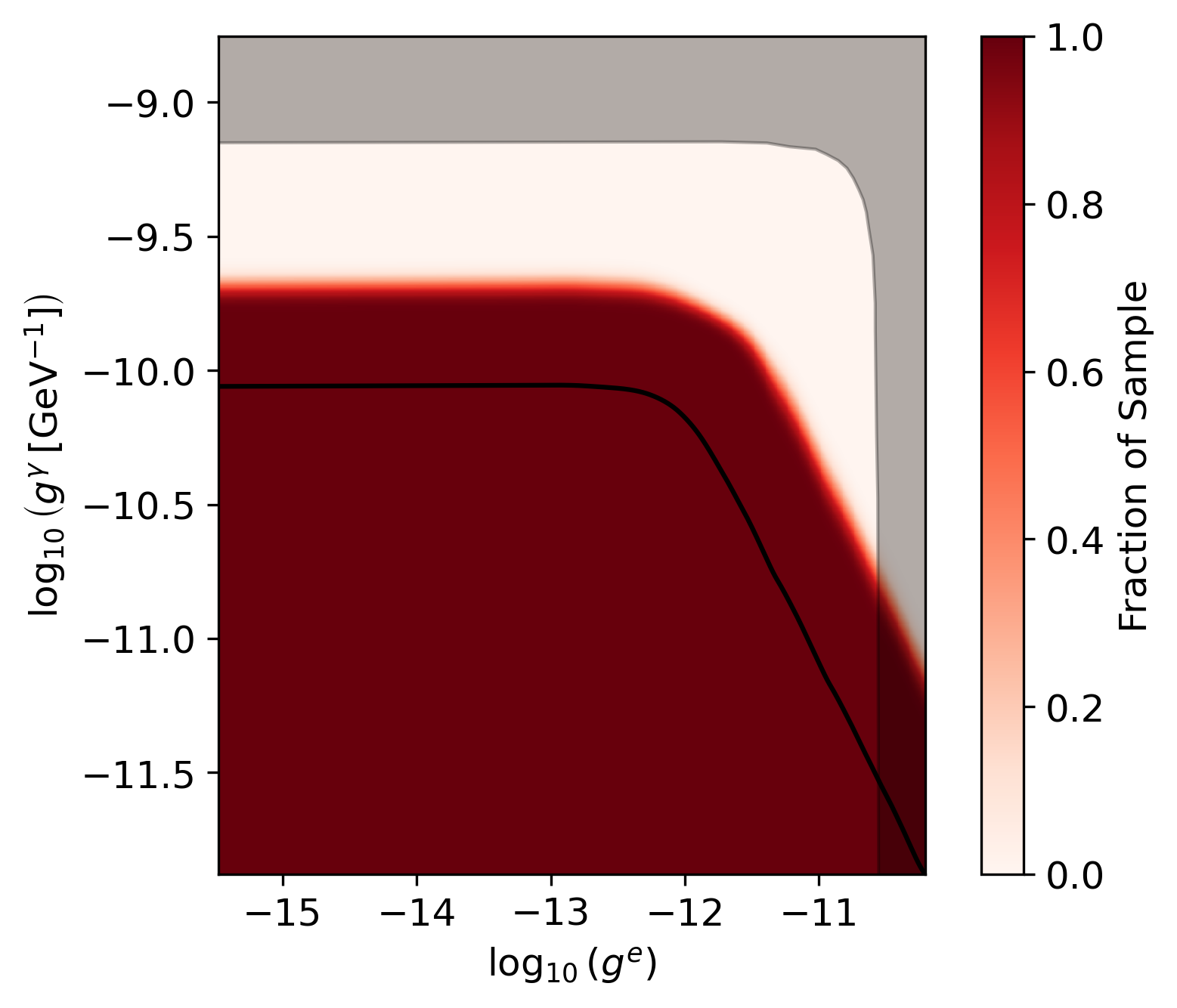}
\centering
\caption{The fraction of realisations consistent with non-detection as a function of coupling ($g^{e}$ and $g^{\gamma}$) shown for two different values of $N$ -- Left: $N=2$; Right: $N=30$. The grey region indicates the excluded region from solar neutrinos \cite{Gondolo:2008dd}.}
\label{fig:reCAST}
\end{figure}
%%%%%%%%%%%%%%
From \figref{fig:reCAST}, it can be seen that increasing the number of ALPs decreases the competitiveness of the bound. 
The left plot of \figref{fig:reCAST} shows the bound with $N=2$ ALP fields, and we observe that the effect of an additional state is marginal; however, for $N=30$,  we observe that the bound on $g^\gamma$ is relaxed by almost half an order of magnitude with $\log_{10}(g^\gamma\,[\rm{GeV^{-1}}]) \lesssim -9.6$. As the number of hidden states increases, the oscillated flux of electromagnetic ALPs on Earth decreases since they can oscillate into the hidden ALP states that are not detectable by CAST. Hence, the effective coupling  $g^\gamma$ can increase to compensate for this decrease in the detectable flux.

To quantify this relationship, we consider ALP multiplicities $N \in \left[ 2, 30 \right]$. For each $N$ in this set, we determine the value of $g^{\gamma}$, $g^{\gamma}_{50}(N)$ for which $50\%$ of mixing realisations satisfying $\Phi_{\gamma}^{\rm{osc}}<\Phi^\mathrm{Max}(g^{e}, g^{\gamma})$ at a point in the horizontal region of (\figref{fig:reCAST}). We fix $g^{e} = 10^{-15}$. For this low ALP-electron coupling, the production processes for $\phi_e$, namely bremsstrahlung and Compton scattering, are ineffective, and the production of $\phi_\gamma$ in the Sun dominates. $g^{\gamma}_{50}(N)$ can be interpreted as an approximate bound on $g^{\gamma}$ in the ALP anarchy scenario. This result is shown in \figref{fig:extrapBounds}, which we fit the following function to:
\begin{equation}
    g^{\gamma} = e^{ m\log N - c},
\end{equation}
where we find $m = 0.25$ and $c = -23$. This fitting function is also shown in \figref{fig:extrapBounds}.
\begin{figure}
  \centering
  \includegraphics[width=0.5\textwidth]{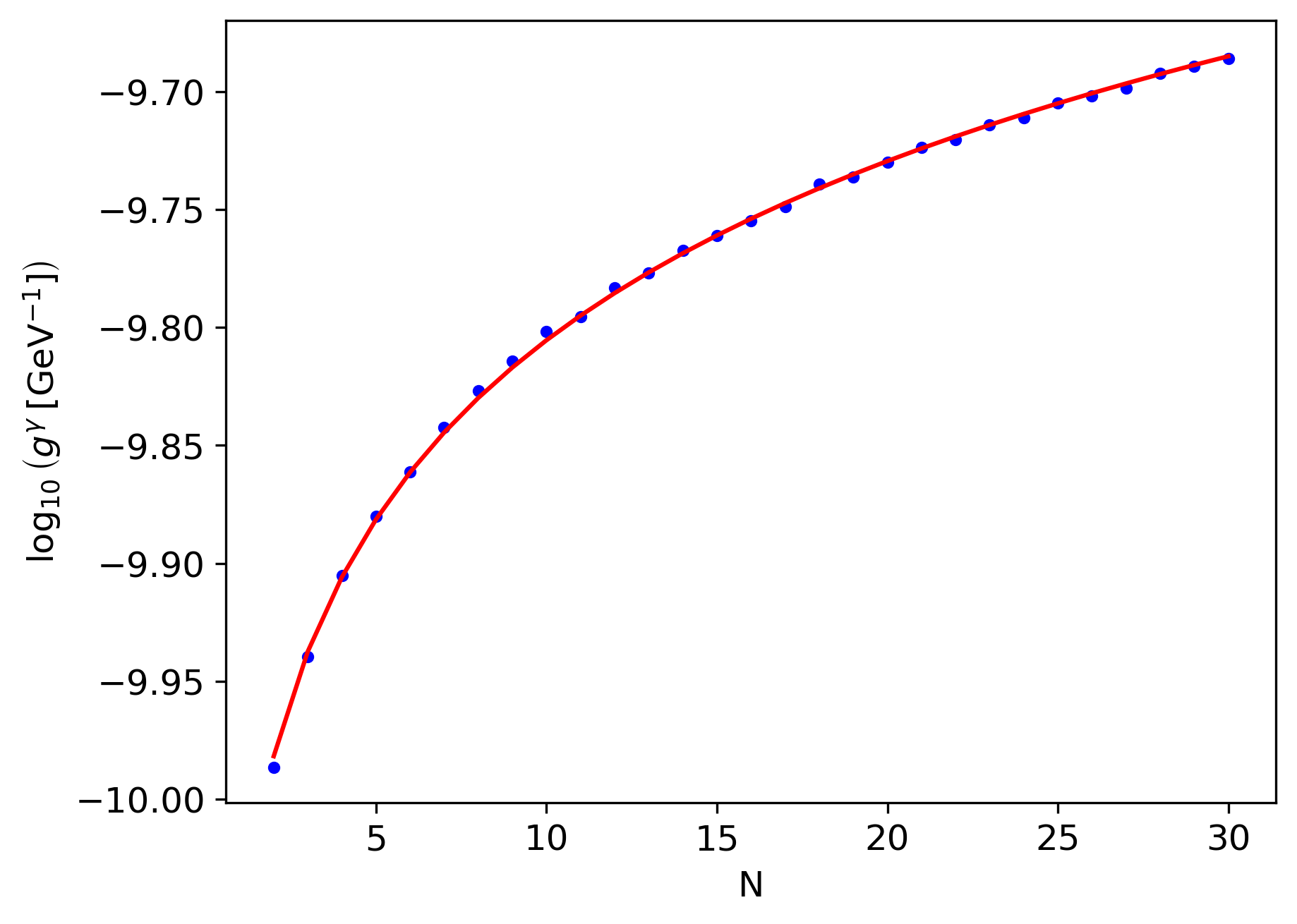}
\centering
\caption{Bounds on photon ALP coupling ($g^{\gamma}$) as a function of number of ALPs ($N$) for $g^{e} = 10^{-15}$. The scatter plot depicts the minimum upper bound consistent with $50\%$ of mixing matrix realisations. The fit to that is shown as a line plot.}
\label{fig:extrapBounds}
\end{figure}
We can understand the dependence of the bound on $N$ -  $g^{\gamma}_{50}(N) \propto N^{1/4}$. We first note that when electromagnetic ALP production dominates, CAST is sensitive to ${g^{\gamma}}^4$, as the ALP must be produced and detected via this coupling to photons. In the many ALP scenario, the bound placed on ${g^{\gamma}}^4$ is weakened in proportion to the electromagnetic ALP survival probability $P_{\gamma \rightarrow \gamma}$ given in \equaref{eqn:gg}. For large $N$, the variance term becomes negligible and we have $P_{\gamma \rightarrow \gamma} \sim \frac{1}{N}$. We therefore obtain $g^{\gamma}_{50}(N) \propto N^{1/4}$, as found numerically.

We have seen that if the ALP-photon and ALP-electron interactions are an effect from multiple ALP mass eigenstates, the CAST bounds on the total effective couplings may be somewhat reduced. We have calculated this reduction in the ALP anarchy scenario for mass differences $\Delta m^2 > 10^{-12} \, \rm{eV}^2$. As shown in \cite{Chadha-Day:2021uyt}, for $\Delta m^2 < 10^{-14} \, \rm{eV}^2$, there is no significant oscillation into hidden states. For intermediate mass differences, there is a non-trivial oscillation structure depending on $\omega$.
%%%%%%%%%%%%%%%%%%%%%%%%%%%%%
\section{Magnetic white dwarfs}
\label{sec:MWD}
%%%%%%%%%%%%%%%%%%%%%%%%%%%%%
Having considered the constraints from the CAST experiment, we now examine how the limits on the single ALP electromagnetic and electronic coupling from observations of magnetic white dwarfs (MWDs) can be applied to our multi-ALP scenario.
% We now consider limits on ALP couplings from observations of magnetic white dwarfs (MWDs).
MWDs produce $\omega \sim$ keV energy ALPs via axio-bremsstrahlung, which can convert to  X-rays in the magnetosphere surrounding the MWD. Subsequently, searches for observable X-ray signals provide one of the most stringent constraints on the ALP parameter space, see e.g. Ref.~\cite{Dessert:2019sgw}. 
More specifically, in the single ALP scenario, where  $N=1$ and $\phi_e\equiv \phi_\gamma$,  the $(g^e, g^\gamma)$ parameter space is constrained 
from the non-observation of astrophysical X-ray emission from the  MWD  \mbox{RE J0317-853}  by the Suzaku telescope \cite{doi:10.1093/pasj/65.4.73}. The flux of ALP-induced X-ray photons on Earth, in the low ALP mass regime, is approximately proportional to $\Phi_{\text{X-ray}}\propto (g^e  g^\gamma)^2$ as the ALP luminosity is proportional to ${g^e}^2$ while the probability the ALP transitions to X-ray photons is proportional to ${g^\gamma}^2$. The non-observation of excess X-rays provides an upper bound on $\Phi_{\text{X-ray}}$ and hence a corresponding bound on $(g^e, g^\gamma)$.

In the multi-ALP case, since the radius of the MWD is relatively small (less than a percent of the Sun's radius \cite{Kulebi:2010pd}), the oscillations between the electronic ALP and hidden states do not have time to develop, assuming  $\Delta m^2\lesssim 4 R =10^{-10}\,\rm{eV}^2$, where $R$ is the MWD's radius. Moreover, we assume that all ALP masses considered are $ m_{i}< 10^{-2}\,\rm{eV}$.
To perform a simple recast of the single to the multi-ALP case, we compute the conversion probability ($P_{e \rightarrow \gamma}$) and scale the $N=1$ bound on $g^e g^{\gamma}$, denoted as $b_{N=1}$, as follows:
\begin{equation}
    b_{N > 1} = b_{N=1} / P_{e \rightarrow \gamma}
\end{equation}
where $b_{N > 1}$ is the new bound on $g^e g^\gamma$ assuming the existence of $N$ ALP states. A decrease in the conversion probability will allow for a greater possible coupling strength. In the MWD setting, in the case where oscillations do not have time to develop, $P_{e \rightarrow \gamma}$ is given by the scalar product of $\phi_e$ and $\phi_{\gamma}$:

\begin{equation}
      P_{e \rightarrow \gamma} = \left| \frac{\sum_i g_i^e g_i^\gamma}{g^e g^\gamma} \right|^2
\end{equation}

Note that this is a different limit to that considered in \secref{sec:CAST} as the distance propagated is much lower. The resulting bound as a function of number of axions is shown in \figref{fig:WMD} for values: $L = 4 \times 10^{-3} $ Solar Radii \cite{Kulebi:2010pd}, $E = 5 \,\mathrm{keV}$ and masses distributed logarithmically in $[10^{-9}, 10^{-6}]\,\mathrm{eV}$.
The conversion probability, accounting for the presence of hidden ALPs (in the simple two-ALP case, see \equaref{eq:Prob}), is given by
\be
P_{e \to \gamma}= \mathrm{cos}^2(\theta^e - \theta^{\gamma})\,,
\ee
which describes the projection of the electronic ALP state onto the EM ALP state. 
In the limit $\theta^e=\theta^\gamma$, naturally, the probability $P_{e \to \gamma}=1$ because the electron and EM ALP states are completely aligned, and the $N=1$ case is recovered. If these angles are very different, then the probability  $P_{e\to\gamma}$ can be significantly smaller and make the bound proportionally weaker. 
\begin{figure}[t!]
\centering
\includegraphics[width=0.5\textwidth]{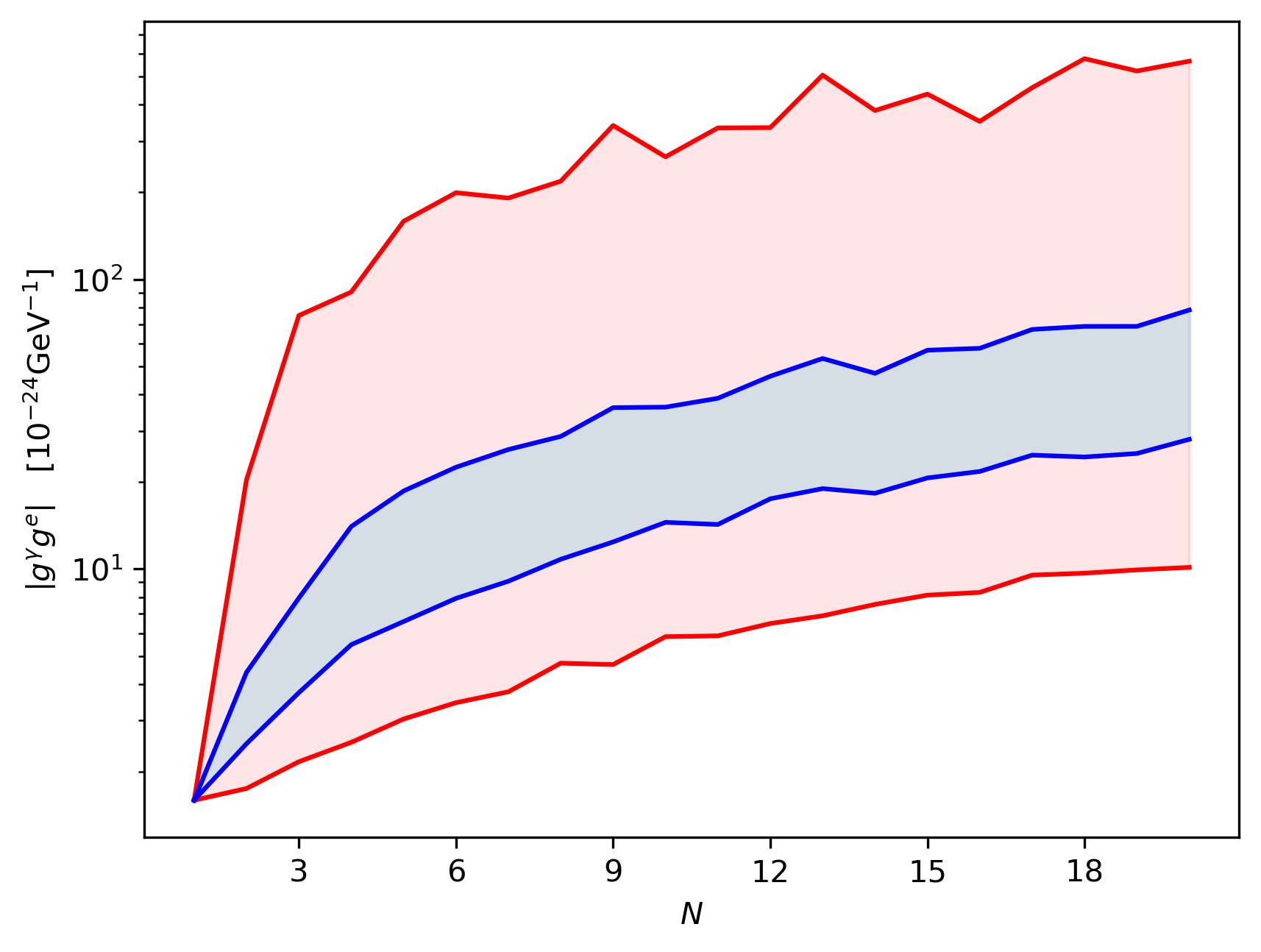}
\caption{The bounds on $g_\gamma g_e$ as a function of $N$ for the low mass ($m_a < 10^{-6} \mathrm{eV}$), low propagation limit of the MWD. The bounds were computed for a set $1000$ coupling pairs ($g^e$ and $g^\gamma$). The central $90\%$ of bounds lie within the red band, with the central third being encompassed by the blue band. The one ALP bound was recast from Ref.~\cite{Dessert:2019sgw}.}\label{fig:WMD}
\end{figure}
In \figref{fig:WMD}, we show how the limit on $\lvert g^\gamma g^e\rvert$, in the low ALP mass regime, from Ref.~\cite{Dessert:2019sgw} changes as the number of hidden states is increased. 
We observe that as the number of hidden states is increased, the total allowed coupling strength of the ALP to EM and electrons increases. This occurs because increasing the number of hidden states leads to an increase in the typical orthogonality between the electromagnetic and electronic ALP states, decreasing the chance that an ALP produced in an electron interaction will convert into a photon. We find that the bound decreases almost three orders of magnitude as the number of ALP mass states increases from $2$ to $18$. 

\section{Very high energy blazars}
\label{sec:blazar}
In this final section, we examine how our multi-ALP scenario can be constrained by the observation of very high-energy gamma rays. We begin by outlining the simulation of the propagation of these high-energy photons emitted by blazars toward Earth, considering the possibility of photon mixing with multiple ALP states. We detail the density matrix equations and the model of the magnetic fields used in the simulation. Additionally, we discuss the selection of blazars and how these sources can be utilised to constrain the electromagnetic coupling as a function of the number of ALPs in our multi-ALP scenario.

Blazars produce a large flux of Very High Energy (VHE) photons. These TeV scale photons can scatter off the 
isotropic extragalactic background light (EBL) as they propagate to Earth, producing positron-electron pairs.
The probability with which this scattering occurs increases with energy. As such, we expect significant attenuation of high-energy photons travelling through intergalactic space. Telescope observations suggest that the Universe may be more transparent than expected to VHE photons \cite{2014JETPL, Kohri_2017}, although the evidence for this effect is not conclusive \cite{Biteau_2015, Dominguez:2015ama}. Several phenomenological studies have introduced ALPs to explain this discrepancy \cite{DeAngelis:2011id,Simet_2008,Sanchez-Conde:2009exi,Meyer_2013}. VHE photons can oscillate into ALPs in the magnetic field of the blazar or the intergalactic medium and thus travel unimpeded through the Universe. These oscillations may conspire for appropriate masses and couplings, such that the ALPs reconvert into photons in the Milky Way, allowing for their detection on Earth. In this case, the measured flux of VHE photons on Earth will be amplified, accounting for the observations. 
In this section, we will explore this effect in the context of our multi-ALP model. Following and extending the analysis of Ref.~\cite{Meyer_2013}, we consider an arbitrarily mass-mixed set of ALP states. We note that the ALP-electron coupling does not play a significant role in this process, and therefore will not be considered in this section.

\subsection{Simulation}
\begin{figure}[t!]
\centering
\includegraphics[width=0.9\textwidth]{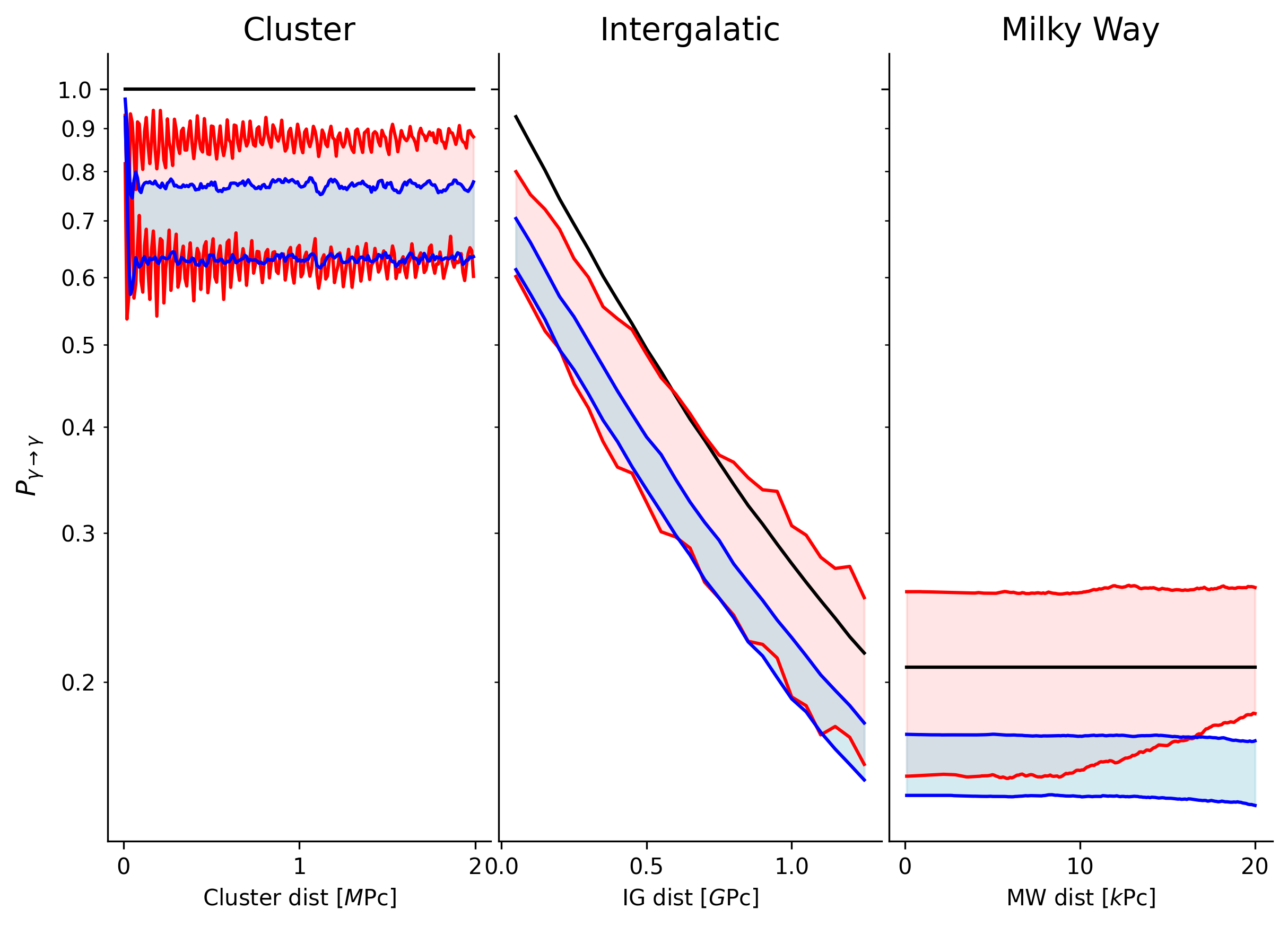}
\caption{Photon survival probability against propagation distance for a photon energy of $400\,\rm{GeV}$ produced by 1ES0414+009. The zero ALP case is shown in black, with the central third of samples shown in red and blue for the 1 ALP and 20 ALP cases respectively.}\label{fig:stepWise}
\end{figure}

% \begin{figure}[t!]
% \centering
% \includegraphics[width=0.7\textwidth]{fracGreater.png}
% \caption{The proportion of magnetic field runs that result in a greater $P_{\gamma \rightarrow \gamma}$ than the zero ALP case as a function of energy for 1 and 20 ALPs. The 1 ALP case is shown with a dashed line and the 20 ALP case, with a dashed dot line. The source 1ES0414+009 was used to produce this figure.}\label{fig:stepWise}
% \end{figure}

%
In the following subsections, we describe the approach by which we simulate the propagation of VHE photons from their blazar source to Earth. We consider both photon-EBL scattering and ALP-photon mixing. As the ALP mass is relevant for this effect, we work in the mass basis where each ALP couples separately to the photon. Photon-EBL scattering is dissipative (VHE photons can scatter off EBL photons, creating positron-electron pairs) and introduces non-unitary into the evolution. To account for this effect, we use a density matrix formalism.

Our simulation can be broken down into three spatial regions: propagation through the galaxy cluster hosting the blazar; propagation through the intergalactic medium (IGM); and propagation through the Milky Way (MW). The principal difference between each region is the form of the magnetic field present. Photon to ALP conversion occurs most readily in the galaxy cluster where the VHE-photon density and magnetic fields are large. As the VHE photon/ALP beam propagates out of the cluster, the relatively strong cluster fields give way to a much weaker intergalactic field, largely suppressing any ALP-photon processes. The vast IGM is instead responsible for the attenuation of photons by EBL scattering. Finally, once the blazar jet reaches the Milky Way, we expect to see ALP-photon back-conversion induced by the strong galactic magnetic fields.

The effect of each propagation region is depicted in \figref{fig:stepWise} which shows the photon survival probability of a photon with $E=400\,\rm{GeV}$ as a function of the distance of propagation from the blazar. The solid black line indicates the scenario without ALPs ($N=0$), while the red and blue shows the effect of $N=1$ and $N=20$ ALP states. In the no-ALP scenario, the survival probability remains unity until propagation through the IGM where, due to EBL scattering, it decreases to $\sim 0.2$. For the cases that include ALPs, a number of model unknowns associated with oscillation lead us to consider a large set possible $P_{\gamma\rightarrow\gamma}$ -- each element generated with a unique B-field structure and, for $N > 1$, an anarchical choice of $\{g^\gamma_i\}$. The central third of these $\{P_{\gamma\rightarrow\gamma}\}$ are indicated by the filled regions in the figure. In both the one and 20 ALP cases, we see the effect of photon to ALP conversion in the galaxy cluster. In the one ALP case, we see a significant increase in the photon survival probability as the beam travels through the Milky Way, corresponding to the reconversion of ALPs to photons. However, this increase is not present in the 20 ALP case. This is because the electromagnetic ALP produced in the galaxy cluster is very likely to oscillate into a hidden ALP before it reaches the Milky Way. Due to oscillation into hidden ALPs, if the ALP-photon coupling is spread over 20 ALPs, we always expect a {\it decrease} in the blazar luminosity rather than the increase seen in the single ALP case.

%The upper limits on $P_{\gamma\rightarrow\gamma}$ correspond to a near-maximal photon state. This is suppressed in the many ALP case, where, under uniform sampling of $\{g^\gamma_i\}$, Photon-ALP conversion events are secondary to inter-ALP oscillations. The lower limits in the figure correspond to a near-maximal ALP state and are largely independent of the number of ALP's in the first two propagation regions. In the MW, however, we see a significant uptick in the lower limit for the 1 ALP case, corresponding to Photon-ALP back conversion induced by the strong magnetic field. In the many ALP case, the plurality of oscillation channels favours inter-ALP oscillation over back-conversion and we do not reproduce the results of the single ALP case. The overall widths of the red and blue regions indicate the degree of uncertainty in $P_{\gamma\rightarrow\gamma}$ introduced by varying the B-field and couplings ($\{g^\gamma_i\}$). Increasing $N$ has the additional effect of reducing this uncertainty as hidden states, having no photon coupling, are inert to B-field variations.

\subsubsection{The galaxy cluster}

The simulation begins with a jet of pure photons at the site of the blazar. These photons are unpolarised, hence we take as our initial state to be the following the density matrix:
\begin{equation}
\rho_\gamma = 
1/2
 \begin{bmatrix}
    1 & 0 & 0 & 0 \\
    0 & 1 & 0 & 0 \\
    0 & 0 & 0 & 0 \\
    0 & 0 & 0 & \ddots\,
\end{bmatrix}\,,\label{eqn:stateMat}
\end{equation}
where the first two diagonal elements correspond to the photons' two polarisations, and the remaining diagonal elements correspond to the ALP mass eigenstates; off-diagonal elements are associated with a superposition of states. To facilitate comparison between the multi-ALP and single ALP effects, we follow \cite{Meyer_2013} in our description of the magnetic fields. We take the cluster field to have a domain-like structure with a radially dependent magnitude given by:
\begin{equation}
    B^C(r) = B_0^C(1+(r/r_\mathrm{core})^2)^{-\eta}\,,
\end{equation}
\noindent
where, the core radius is $r_\mathrm{core} = 200\,\mathrm{kpc}$ and the central field strength is $B_0^C = 10\,\mu \mathrm{G}$ and $\eta =0.5$. Within a given magnetic domain, the field is assumed to be constant. The direction of the magnetic field is randomised in each domain. The coherence of the magnetic field, therefore, depends on the domain length, which is taken to be $\Delta L_\mathrm{c} =10\, \mathrm{kpc}$. The total radius of the cluster is $2\, \mathrm{Mpc}$.

Galaxy clusters are home to a large population of charged particles that will give an effective mass to the photon. To accurately describe these, we consider the electron density given by:
\begin{equation}
    n_\mathrm{el} (r) = n_\mathrm{el}^0 (1 + r/r_\mathrm{core})^{-1}\,,
\end{equation}
\noindent
where $n_{el}^0 = 10^{-2} \mathrm{cm}^{-3}$.
With this description of the cluster, we evolve the state matrix of \equaref{eqn:stateMat} by constructing an evolution operator ($G$) for each domain:
\be
\ba
    G &= e^{i H \Delta L}\,, \label{eqn:G}\quad 
    H =  \begin{bmatrix}
    \Delta^{\mathrm{pl}}+2\Delta^{\mathrm{QED}} & 0 &(\Delta^{\phi\gamma}_x)_1 & (\Delta^{\phi\gamma}_x)_2 & (\Delta^{\phi\gamma}_x)_3 \\
    0 & \Delta^{\mathrm{pl}}+\frac{7}{2}\Delta^{\mathrm{QED}} & (\Delta^{\phi\gamma}_y)_1 & (\Delta^{\phi\gamma}_y)_2 & (\Delta^{\phi\gamma}_y)_3 \\
    (\Delta^{\phi\gamma}_x)_1 & (\Delta^{\phi\gamma}_y)_1 & \Delta_1^{\phi} & 0 & 0 \\
    (\Delta^{\phi\gamma}_x)_2 & (\Delta^{\phi\gamma}_y)_2 & 0 & \Delta_2^{\phi} & 0 \\
    (\Delta^{\phi\gamma}_x)_3 & (\Delta^{\phi\gamma}_y)_3 & 0 & 0 & \ddots \\
\end{bmatrix}\,, 
\ea\ee
where $H$ denotes the propagation Hamiltonian with the following parameters:
\be
\ba
\Delta^{\mathrm{pl}} & = \frac{-1.1 \times 10^{-10} \rm{GeV}}{E\times 10^{-3}} \frac{n_e}{10^{-3} c\rm{m}^3}\\
\Delta^{\mathrm{QED}} & = 4.1 \times \frac{10^{-6} \rm{GeV}}{E\times 10^{-3}} \frac{B_x^2 + B_y^2}{\mu\rm{G}^2}\\
\Delta^{\phi} _i & = \frac{-7.8 \times 10^{-3} \rm{GeV}}{E \times 10^{-3}} \left(\frac{m_i}{10^{-8} \rm{GeV}}\right)^2\\
(\Delta^{\phi\gamma}_x)_i & = 7.6\times 10^{-2} \frac{g^{\gamma}_{i} \rm{GeV}}{5 \times 10^{-11}} \frac{B_y}{\mu\rm{G}}\\
(\Delta^{a\gamma}_y)_i & = 7.6\times 10^{-2} \frac{g^{\gamma}_{i} \rm{GeV}}{5 \times 10^{-11}} \frac{B_x}{\mu\rm{G}}\,,\\
\ea
\ee

here, $E$ is the photon/ALP energy (in $\rm{GeV}$), $n_e$ is the electron density and $B_x$ and $B_y$ are the $x$ and $y$ components of the magnetic field strength (in $\mu \rm{G}$) \cite{Masaki_2017}. The mass of the $i^{\rm{th}}$ ALP state, denoted $m_i$, is taken to be distributed logarithmically within $\left[10^{-8}, 10^{-5}\right] \rm{eV}$. The various components of $H$ can be interpreted as follows: $\Delta^{\phi}_i$ are the ALP mass terms; $\Delta^{pl}$ is a photon effective mass terms, induced by thermal effects in the electron plasma; $(\Delta^{\phi \gamma}_x)_i$ and $(\Delta^{\phi\gamma}_y)_i$ are the ALP-photon couplings for each photon polarisation; and, $\Delta^{\rm{QED}}$ implements vacuum polarisation effects. As we are working in the mass basis, each ALP is independent of all other ALP states so $H$ is diagonal in the ALP sector. We construct a new evolution operator, $G_i$, for each domain using the central values of $B$ and $n_\mathrm{el}$. The state of the system after propagation through the cluster is then given by:

\be
\ba
\rho_\mathrm{out}^C =& \tilde{G}^C \rho_\gamma \left(\tilde{G}^C\right)^\dagger\,, \\
\tilde{G}^C =& \prod_i G_i^C\,,
\ea
\ee
where $i$ runs over each domain in the cluster.
\subsubsection{Intergalactic space}

The intergalactic magnetic field is modelled with a similar domain-like structure to that of the AGN cluster, albeit with a significantly lower overall strength and a much larger domain length $\Delta L^{IG} = 50\, \mathrm{Mpc}$. The intergalactic medium has a luminosity redshift ($z$) dependent field strength given by:

\begin{equation}
    B^{IG}(z) = B_0^{IG} (1 + z)^2\,,
\end{equation}
where $B_0^{IG} = 1 \, \mathrm{nG}$. The intergalactic electron density is approximated with a constant value of $n_\mathrm{el}^{IG} = 10^{-7}\,\mathrm{cm}^{-3}$.
To account for VHE photon scattering off the EBL we introduce a non-unitary decay matrix ($D$) for each domain:
\begin{equation}
    D(\tau) = \begin{bmatrix}
    \exp (- \tau / 2) & 0 & 0 \\
    0 & \exp (- \tau /2) & 0 \\
    0 & 0 & \ddots
\end{bmatrix}\,,
\end{equation}
\noindent
where $\tau$ is the optical depth associated with propagation through a given domain. In general, $\tau$ depends on the photon energy and the redshift of the domain. We use the EBL opacity model given in Ref. \cite{Franceschini_2008}. After computing $D(\tau_i)$ for each domain, the ALP-photon state is propagated as follows:
\be
\ba
    \rho_\mathrm{out}^{IG} & = \tilde{G}^{IG}  \rho_\mathrm{out}^C \left( \tilde{G}^{IG} \right)^\dagger , \,\\
    \tilde{G}^{IG} & = \prod_i D(\tau_i) G_i^{IG}\,,
    \ea
\ee
where $G_i^{IG}$ is calculated as before using \equaref{eqn:G} with $i$ running over each domain in the IGM.
\subsubsection{The Milky Way}
Compared to the previous two field scenarios, the Milky Way field has a significantly more complex structure that comprises a halo field that surrounds a spatially compact disk field. Following Ref. \cite{Jansson_2012}, we describe the disk by a series of logarithmic spirals and the halo field by a superposition of piece-wise functions extending relatively far above and below the galactic plane, see \figref{fig:MWField}. It is this extensive halo field that is responsible for the majority of ALP-Photon conversion.
A detailed description of the field model used is provided in \cite{Jansson_2012}. We note that in our work, a small alteration has been made to correct the function describing boundaries between the consecutive log-spiral regions in the disk:
\be
    r_\mathrm{bound} = r_j e^{(\theta-\pi) \cot \theta }\,,\quad\quad 
    \theta = \frac{\pi}{180}(90 - \alpha_\mathrm{open})\,,
\ee
where $\alpha_\mathrm{open} = 11.5^o$ is the opening angle of the log-spiral and $r_j$ are the radii at which each spiral boundary crosses the negative x-axis.
Finally, we use a constant electron density of $0.1 \,\mathrm{cm}^{-3}$ to construct the MW evolution operator. The final state of the system is then given by the product over domains $i$:
\be
\ba
    \rho_\mathrm{out}^{MW} &=   \tilde{G}^{MW} \rho_\mathrm{out}^{IG} \left(\tilde{G}^{MW}\right)^\dagger\,, \\
    \tilde{G}^{MW} &= \prod_i G_i^{MW}\,,
\ea\ee

\noindent
where again we discretise our computation of the evolution operator ($G_i$) - in this case, using an approximate-coherence length of $100\,\mathrm{pc}$. We begin the Milky Way propagation stage when the ALP/photon reaches a radial distance of $20\, \mathrm{kpc}$ from the galactic centre.
The probability that an emitted photon is detected as a photon on Earth is found by projecting the final state  ($\rho_\mathrm{out}^{MW}$) onto the two possible photon states:
\be\ba
    P_{\gamma \rightarrow \gamma} &= \Tr \left( 2\rho_\gamma \mathcal{T}_\mathrm{out}^{MW} \right) \\
    &= \Tr \left( 2\rho_\gamma \left[ \tilde{G}^{MW} \tilde{G}^{IG} \tilde{G}^C\right] \rho_\gamma \left[ \tilde{G}^{MW} \tilde{G}^{IG} \tilde{G}^C\right]^\dagger\right)\,.
\ea\ee

\begin{figure}
  \centering
  \includegraphics[width=0.45\textwidth]{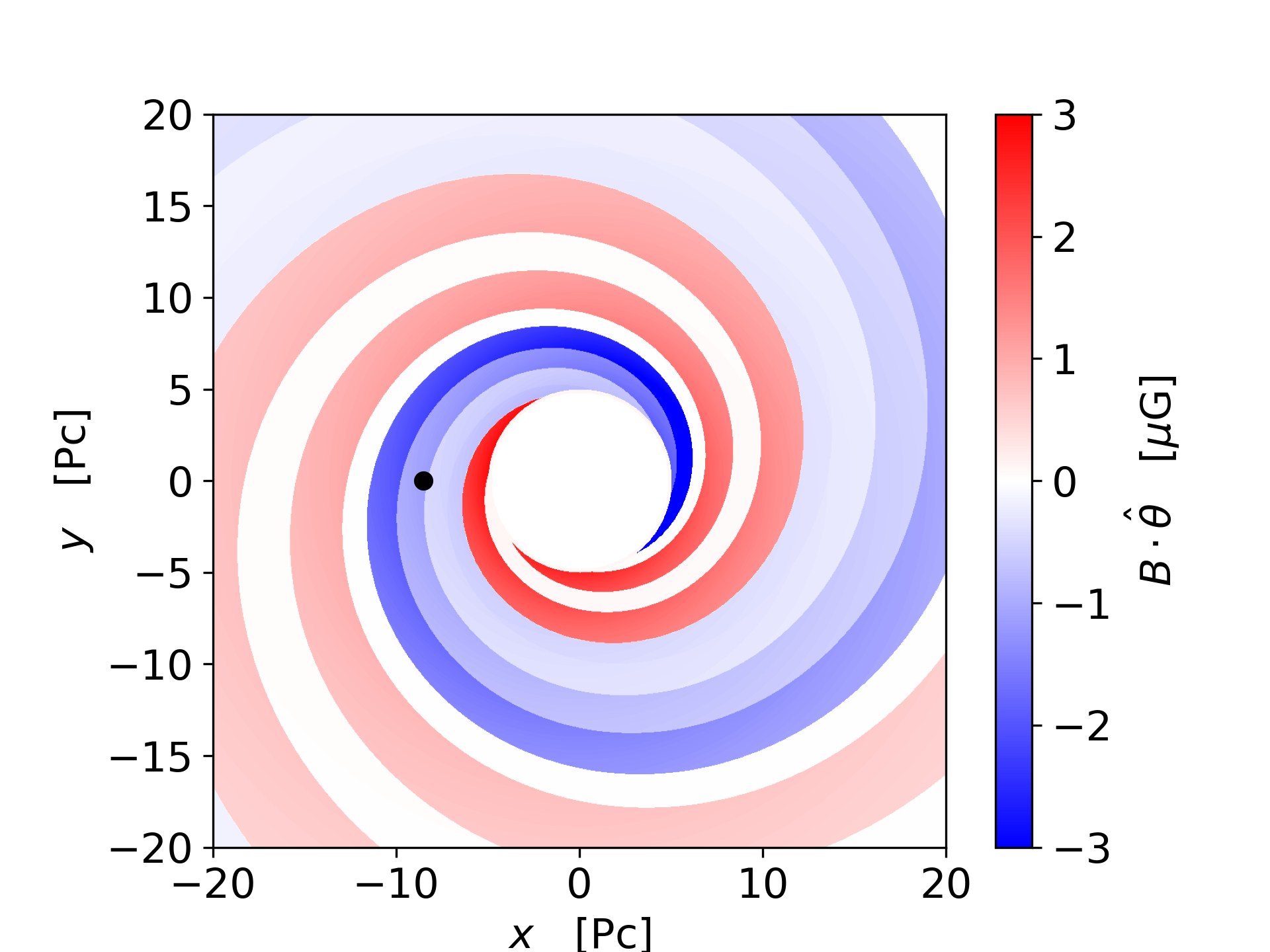}
  \includegraphics[width=0.45\textwidth]{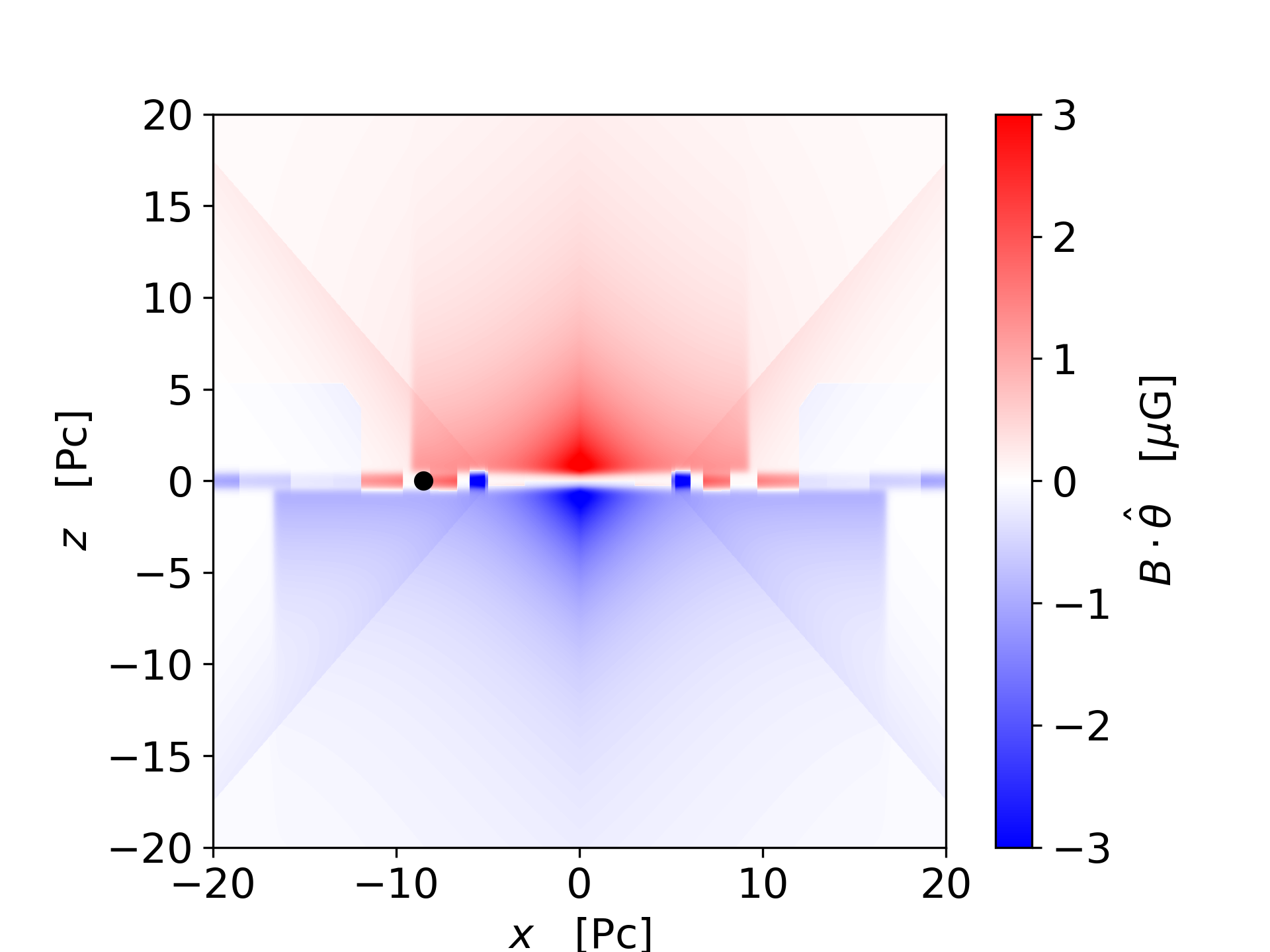}
\centering
\caption{The magnitude of the Milky Way field in two perpendicular planes through the galactic centre. Left: cross-section at $z=0$ looking upon the galactic disk; Right: cross-section perpendicular to disk at $y=0$. The sign of the field indicates the orientation of the azimuthal component: positive - anticlockwise (looking down negative z); negative - clockwise. The location of the earth, $(-8.5, 0, 0) \mathrm{kpc}$, is marked by the black circle.}
\label{fig:MWField}
\end{figure}

\subsection{Results}\label{sec:Results}

Due to observational limitations, the field orientations in each domain of the galaxy cluster and intergalactic medium are unknown and small changes thereto could have a large effect on the final value of $P_{\gamma \rightarrow \gamma}$. It is thus necessary to simulate with a large set of different field directions. $P_{\gamma \rightarrow \gamma}$ now represents a stochastic quantity that we use to marginalise over the unknown field states.

As with the rest of this work, we also encounter an ambiguity in the choice of mixing parameters ($\{g^\gamma_i\}$), which we again account for using the ALP anarchy model described above. Each simulation run has a unique set of coupling parameters and magnetic field orientations. The final results described below were obtained using a sample of $N_\mathrm{samp} = 1000$ simulation runs and a set of $6$ Blazar sources. These sources (listed in \tabref{tab:spectratab}) were selected based on the availability of their data and on their inclusion in other works on this subject \cite{Meyer_2013}. We plan to perform a more detailed follow-up study using a larger sample of sources and more recent data.

\begin{table}[t!]
\centering
\begin{tabular}{|c|l|l|l|}
\hline
$j$ & Source & Experiment & Fit Function \\
\hline
1 & Mkn 501 & HEGRA & LP \\
2 & 1ES0414+009& H.E.S.S & PL \\
3 & 1ES0229-200 & H.E.S.S. & PL \\
4 & Mkn421 & H.E.S.S. & LP \\
5 & 1ES1101-232 & H.E.S.S & PL \\
6 & 1ES0347-121 & H.E.S.S & PL \\
\hline
\end{tabular}
\caption{List of  VHE gamma-ray sources used in our analysis and their corresponding fit functions; Log Parabola (LP) or Power Law (PL), see \appref{sec:AppB}.}
\label{tab:spectratab}
\end{table}
For a given choice of blazar source and grid point ($N, g^\gamma$) in model parameter space, the result of our simulation is a set of $N_\mathrm{samp}$ survival probability spectra $P_{\gamma\rightarrow\gamma}(E)$, where $E$ is the photon energy. To marginalise over these, we use the students p-statistic ($p_t$), following \cite{Meyer_2013}; the calculation of $p_t$ is described in \appref{sec:AppB}. For each grid point, we obtain a distribution of $N_\mathrm{samp}$ p-statistics. We collate these data by determining the value of $p_t$ corresponding to the field configuration, resulting in a better agreement between the model and corrected spectra than $95\%$ of other configurations. We denote this as $p_{95}$. In this way, we arrive at a two-dimensional parameter scan for each source that preserves more of the underlying $p_t$ distribution than could be achieved, for example, by simply averaging the $p_t$ over $N_\mathrm{samp}$. This method facilitates comparison with Ref. \cite{Meyer_2013}. Note that this method results in relatively conservative bounds on ALP scenarios as we are choosing a B field configuration that agrees rather well with the data. 

Finally, to construct an overall parameter scan we determine $p_{95}$ on the union of the sample data for each source. The resulting grid scan is shown as a heat map in \figref{fig:BlazarFinal}. Lower values of $-\log_{10} \left( p_{95} \right)$ correspond to the model better reproducing the observations. We see a significant improvement over the Standard Model with the addition of a single ALP state, this corroborates the findings of Ref. \cite{Meyer_2013}; the introduction of ALP states can alleviate tensions in standard intergalactic opacity models. Interestingly, and novel to this work, we also observe that increasing $N$  provides a worse fit to the VHE data. As the number of ALP states increases, so too does the number of hidden states. Working in the interaction basis, states in models with many ALPs will exist less frequently as the interaction state. We might expect, therefore, a suppression in the back conversion probability. This can be seen explicitly in \figref{fig:stepWise}, where in the $20$ ALP case, we do not observe any increase in $P_{\gamma \rightarrow \gamma}$ in the MW. This effect can be very significant, in this case, completely negating the opacity decrease caused by reduced EBL scattering; $P_{\gamma \rightarrow \gamma}$ is significantly lower in the $20$ ALP case than it is for no ALPs (black line in the \figref{fig:stepWise}). It should be noted that the results depicted in \figref{fig:stepWise}, having been computed for a single source and energy, may not be representative. In contrast, \figref{fig:BlazarFinal} reflects our statistical analysis using multiple blazars and energies, where the single ALP state shows the greatest agreement with observed spectra.
In addition to favouring fewer ALP states, we also see a lower degree of tension at greater ALP couplings. The lower regions of \figref{fig:BlazarFinal} behave as no ALP models, and there is no significant reduction in EBL scattering.

\begin{figure}[t!]
\centering
\includegraphics[width=0.8\textwidth]{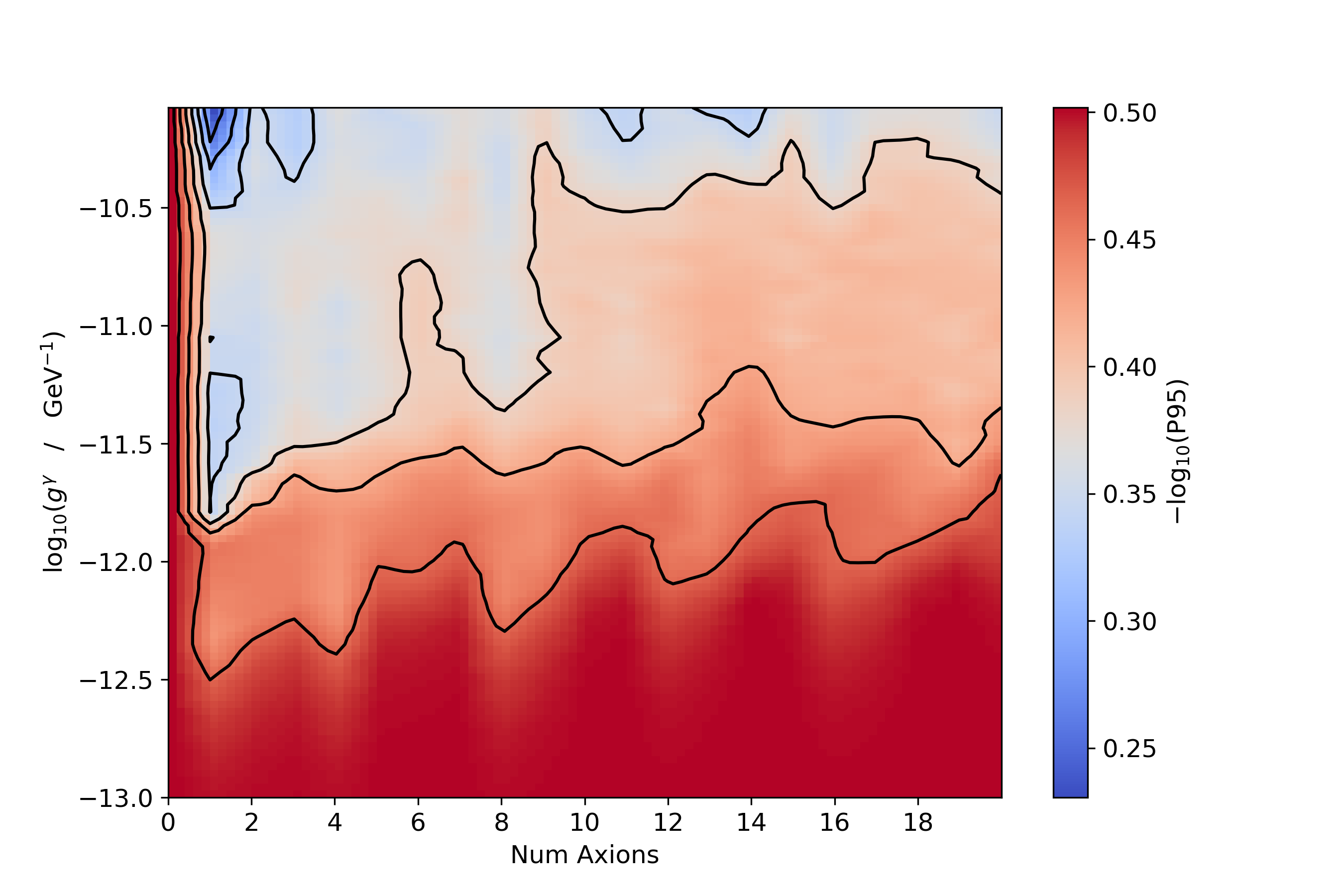}
\caption{A heat map showing $p_{95}$ computed using $6$ sources and $N_\mathrm{samp} = 1000$ samples per source. Lower values (higher $p_{95}$) indicate better agreement between the model and the data.}\label{fig:BlazarFinal}
\end{figure}
%%%%%%%%%%%%%%%%%%%%%%%

%\textbf{Summary}
%Note that two statistical tests are performed: one to obtain the power law fit to parametrise the no new physics scenario and a second to parametrise how compatible the new physics is with this flux given the uncertainties in the data. 

\section{Discussion and conclusions}
\label{sec:conclusion}
In this paper, we have explored the phenomenology of ALP anarchy models comprising many axion-like particles whose masses and Standard Model couplings are related by random matrices. String compactifications typically generate many ALPs; therefore, the phenomenology of many ALP systems is an important direction in studying physics beyond the Standard Model. The ALP anarchy scenario provides a benchmark for this phenomenology.

We have shown that a key feature of many ALP phenomenology is oscillations between the ALP states that couple to the Standard Model and hidden ALP states that do not. A given ALP model can be characterised by the total ALP photon and ALP electron couplings ${g^{\gamma}}^2 = \sum {g_i^{\gamma}}^2$ and ${g^{e}}^2 = \sum {g_i^{e}}^2$. However, oscillation into hidden ALP states can significantly reduce the signal in some ALP searches, such as CAST and IAXO. Other ALP search strategies sensitive only to photon disappearance into ALP degrees of freedom, such as arguments from stellar cooling, will be largely insensitive to the number of ALP fields for a given total ALP-photon coupling. Still, other search strategies such as ADMX \cite{ADMX:2018ogs} and other Dark Matter haloscopes rely on a mass resonance and would, therefore, be sensitive to each ALP mass eigenstate individually rather than to the total effective ALP couplings. If the total ALP-photon coupling and dark matter density is shared over many ALP states, the expected signal in haloscope experiments for a given mass would be correspondingly reduced.

As discussed in section \secref{sec:blazar}, ALPs have been proposed as a solution to increase the observed transparency of the Universe to very high energy photons. However, we found that for many ALP systems, the photon survival probability instead decreases due to the presence of ALPs - the opposite effect to that observed for a single ALP. We, therefore, conclude that many ALP phenomenology leads to a number of new effects not captured by consideration of a single ALP field.

\section*{Acknowledgements}
We thank Anthony Brown, Christopher Dessert, James Halverson, Sebastian Hoof and David Marsh for valuable conversations. We further thank Sebastian Hoof for careful reading and comments on the paper. This work used the DiRAC@Durham facility managed by the Institute for Computational Cosmology on behalf of the STFC DiRAC HPC Facility (www.dirac.ac.uk). The equipment was funded by BEIS capital funding via STFC capital grants ST/P002293/1, ST/R002371/1 and ST/S002502/1, Durham University and STFC operations grant ST/R000832/1. DiRAC is part of the National e-Infrastructure. 
This work was partly performed in part at the Aspen Center for Physics, which the National Science Foundation supports grant PHY-2210452. The authors are supported by STFC grant ST/T001011/1. FCD is supported by EPSRC Stephen Hawking Fellowship EP/T01668X/1.

\section*{Data Access Statement}

The code used to simulate VHE photon propagation from blazars is hosted in the following repository:
\url{https://gitlab.dur.scotgrid.ac.uk/James_Maxwell/blazarphotonaxion}

\appendix
\section{Appendix A: Statistical approach to producing mixing matrices using the Haar measure}\label{sec:AppA}
Here we can outline how the mixing matrices are inductively produced. Following \secref{sec:Anarchy}, one can generate a sample of $N-1$ mixing angles, $\{ \theta_{ij}\}$, in spherical polar coordinates. As a simplifying assumption, we take the mixing matrices, $U\in SO(N)$. Since $SO(N)$ is parametrised by  $N(N-1)/2$ mixing angles, $\left\{\theta_{i j}\right\}_{1 \leq i \leq j \leq N-1}$, the mixing matrix for  $SO(2)$ can  be parametrised by a single angle, $\theta_{11}$:
\begin{equation}
    U_2 =
    \begin{bmatrix}
        \cos \theta_{11} & \sin \theta_{11} \\
        -\sin \theta_{11} & \cos \theta_{11}
    \end{bmatrix}\,.
\end{equation}
The Haar measure of this matrix is given by \equaref{eq:Haar} and explicitly is
\be
dV = d\theta_{11}\,,
\ee
which informs us we sample over $\theta_{11}$ uniformly in $[0, 2\pi]$. Generalising to higher dimensions, the structure of the matrices becomes much less trivial. The most straightforward approach to writing down a higher $SO(N)$ matrix is as a product of matrices:
\begin{equation}
    U_N = U_N' S_N(U_{N-1})
\end{equation}
where $S_N(U_{N-1})$, given by:
\begin{equation}
\ba
    S_N(U_{N-1}) = 
    \begin{bmatrix}
        U_{N-1} & \boldsymbol{0} \\
        \boldsymbol{0}^T & 1
    \end{bmatrix}\,,
    \ea
\end{equation}
\noindent
where $\boldsymbol{0}$ is a $(N-1)$ dimensional zero vector and $S_N(U_{N-1})$ represents rotations in the hyperplane orthogonal to the $N$th dimension. $U_N'$ is the $N$-dimensional analog of $U_2$, describing rotations that involve the $N$th coordinate:
\be
\ba
    {U_N^\prime}_{1, 1} &= \cos \left( \theta_{1, N-1} \right) \\
   U_{N 1, i+1}^{\prime} & =\cos \left(\theta_{i+1, N-1}\right) \prod_{m=1}^i \sin \left(\theta_{m, N-1}\right) \quad \forall \quad i \in[1, N-2] \\
     {U_N^\prime}_{1, N} &= \prod_{m=1}^{N-1} \sin \left( \theta_{m, N-1} \right)\\
    {U_N^\prime}_{i, j} &= 0 \quad \forall \quad j<i-1 \And i > 1 \\
    {U_N^\prime}_{i, j} &= - \sin \left( \theta_{j, N-1} \right) \quad \forall \quad j=i-1 \And i > 1\\
    {U_N^\prime}_{i+1, i+1} &= 
\cos \left(\theta_{i, N-1}\right) \cos \left(\theta_{i+1, N-1}\right) \quad \forall \quad i \in[1, N-2]  \\
     {U_N^\prime}_{N, N} &= \cos \left( \theta_{N-1, N-1} \right) 
\ea
\ee
Explicitly, the $SO(3)$ mixing matrix has the following form:
\be
U_3=\left[\begin{array}{ccc}
\cos \theta_{12} & \sin \theta_{12} \cos \theta_{22} & \sin \theta_{12} \sin \theta_{22} \\
-\sin \theta_{12} & \cos \theta_{12} \cos \theta_{22} & \cos \theta_{12} \sin \theta_{22} \\
0 & -\sin \theta_{22} & \cos \theta_{22}
\end{array}\right]\left[\begin{array}{ccc}
\cos \theta_{11} & \sin \theta_{11} & 0 \\
-\sin \theta_{11} & \cos \theta_{11} & 0 \\
0 & 0 & 1
\end{array}\right]\,.
\ee
Hence we produce the symbolic form of the N-dimensional mixing matrices inductively.
We then sample the angles according to the Haar measure as shown in \equaref{eq:Haar}, and we can visualise these samples as populating an $N$-dimensional sphere uniformly. 

\section{Producing CAST bounds}\label{sec:CASTBounds}
This section provides a detailed overview of the approach used to recast the CAST bounds in  \secref{sec:CASTRes}.  The original CAST bound can be generalised by considering the detector's sensitivity. From any point $(g^{ \gamma}_{N=1}, g^{e}_{N=1})$ on the original bound, the black line shown in \figref{fig:reCAST}, we can extract a proxy for the maximum signal ($\sigma^\mathrm{max}$)  that is consistent with non-detection: 
\begin{equation}
    \sigma_\mathrm{max} = \Phi_{\rm tot}(g^{e}_{N=1}, g^{\gamma}_{N=1}) (g^{\gamma}_{N=1})^2\,,
\end{equation}
where $\Phi_{\rm tot}$ is the total single ALP flux at the detector; computed using \equassref{eq:intB}{eq:intC}{eq:intP}:
\begin{equation}
    \Phi_{\rm tot}(g^{e}_{N=1}, g^{\gamma}_{N=1}) = \Phi_P(g^{\gamma}_{N=1}) + \Phi_B(g^{e}_{N=1}) + \Phi_C(g^{e}_{N=1})\,.
\end{equation}
\noindent
For any point in the coupling parameter space, the total flux at the detector must, therefore, be less than:
\begin{equation}
    \Phi_{\mathrm{max}}(g^{\gamma}) = \sigma_\mathrm{max} / {g^{\gamma}}^2 \label{eqn:totMaxF}
\end{equation}

\noindent
We note that to obtain the total flux at the detector we integrated over the differential fluxes of \equassref{eq:diffintB}{eq:diffintC}{eq:diffintP} using a lower integration boundary of $\omega = 1\,\rm{keV}$ (corresponding to the detector threshold) and an upper bound of $\omega = 10\,{\rm keV}$.

The recast bounds in \figref{fig:reCAST} are shown as heat maps. The weight assigned to each set of couplings $(g^{e}, g^{\gamma})$, corresponds to the proportion of our sample of survival and conversion probabilities, $\{P_{\gamma \rightarrow \gamma}\}$ and $\{P_{e \rightarrow \gamma}\}$, that are consistent with \equaref{eqn:totMaxF}. That is to say that, for a given $(g^{e}, g^{\gamma})$ and probability $P_{e \rightarrow \gamma}^l\in \{ P_{e \rightarrow \gamma} \}$, we find the proportion ($\eta$) of $\{P_{\gamma \rightarrow \gamma}\}$ that satisfy:
\begin{equation}
    \Phi_{\mathrm{max}}(g^{\gamma}) \geq \Phi_P(g^{\gamma}) P_{\gamma \rightarrow \gamma} + (\Phi_B(g^{e}) + \Phi_C(g^{e})) P_{e \rightarrow \gamma}
\label{eq:maxFlux}
\end{equation}
\noindent
At this point, $\eta$ could be found by Monte Carlo sampling $\{ P_{\gamma \rightarrow \gamma} \}$ and determining the number of elements that satisfy \equaref{eq:maxFlux}. This approach is, however, very computationally expensive. A significant saving can be achieved by instead considering the critical value ($P_{\mathrm{Crit}}$) of $P_{\gamma \gamma}$ that achieves equality in \equaref{eq:maxFlux}:
\begin{equation}
    P_{\mathrm{Crit}}(g^{e}, g^{\gamma}, P_{e \gamma}^l) = \frac{1}{\Phi_P(g^{\gamma})}\left[\Phi_\mathrm{max}(g^{\gamma}) - (\Phi_B(g^{e}) + \Phi_C(g^{e})) P_{e \rightarrow \gamma}^l\right]\,,
\end{equation}
\noindent
$\eta$ is now given by the cumulative sum -- evaluated at $P_{\mathrm{Crit}}$ --  of the histogram of $\{ P_{\gamma \rightarrow \gamma} \}$; $\eta = C(P_{\mathrm{Crit}})$. Repeating this process for every $P_{e \rightarrow \gamma}^l\in \{ P_{e \rightarrow \gamma} \}$, the total weight associated with a coupling pair is given by:
\begin{equation}
    W(g^{e}, g^{\gamma}) = \frac{1}{N_{ae}} \sum_{l} C(P_{\mathrm{Crit}}(g^{e}, g^{\gamma}, P_{e \rightarrow \gamma}^l))\,,
\end{equation}
\noindent
where $N_{ae}$ is the size of $\{ P_{e \rightarrow \gamma} \}$ -- taken here to be $N_{ae} = N_{a\gamma} = 10^4$. $W(g^{e}, g^{\gamma})$ is equivalent to the fraction of $\{P_{\gamma \rightarrow \gamma}\}$ and $\{P_{e \rightarrow \gamma}\}$ that satisfy \equaref{eq:maxFlux} and will saturate to $1$ for sufficiently low couplings.

%%%%%%%%%%%%%%%%%%%%%%%%%%%%
\section{Fit quality for VHE spectra and statistical approach to multi-ALP parameter space}\label{sec:AppB}
%%%%%%%%%%%%%%%%%%%%%%%%%%%%
To assess the fit of our multi-ALP model to the VHE Blazar spectra, we employ the same statistical methods as Refs. \cite{Meyer_2013,Meyer:2013mma}. In particular, we use the observed spectra, $\Phi_i^{\text {obs }}$, shown in \tabref{tab:spectratab}.
%%%%%%%%%%%%%%%%%%%%%%%%%%%%

We note that $\Phi_i^{\text {obs }}$ contains data on the blazar source plus all ambient backgrounds.
We compute the photon survival probability for a given point in the theory parameter space:
\begin{equation}
\left\langle P_{\gamma \rightarrow \gamma}\right\rangle_i(g^\gamma, g^e)=\frac{1}{\Delta E_i} \int_{\Delta E_i} \mathrm{~d} E P_{\gamma \rightarrow \gamma}(E)\,,
\end{equation}
where $E$ is the photon energy, and this probability can be computed on a bin-by-bin basis. The absorption-corrected flux is given by
\begin{equation}
\Phi_i(g^\gamma, g^e)=\left\langle P_{\gamma \rightarrow \gamma}\right\rangle_i^{-1}(g^\gamma, g^e) \Phi_i^{\mathrm{obs}}\,,
\end{equation}
where again, we have kept the theory parameter dependence explicit. For a given point in the model parameter space, $(g^\gamma, g^e)$, we are tasked with understanding its compatibility with the observed spectra. For this purpose, we must quantify the observed flux in the absence of new physics and do so by fitting $\Phi_i(g^\gamma, g^e)$ to a power law:
\begin{equation}
f(E)= \begin{cases}N_0\left(E / E_0\right)^{-\Gamma}, & p_{\text {fit }} \geqslant 0.05 \\ N_0\left(E / E_0\right)^{-\left(\Gamma+\beta_c \ln \left(E / E_0\right)\right)}, & \text { otherwise }\end{cases}
\end{equation}
where $p_{\text {fit }}$ denotes the fit probability. The first power law contains three parameters: $N_0$, $E_0$, and $\Gamma$ while in the second (which is a parabola in log-log space), there are four fit parameters, $N_0$, $E_0$, $\beta_c$ and $\Gamma$. The power law function is used unless the baseline (no ALP) fit probability (as calculated by Ref. \cite{Meyer_2013}) is less than 0.05.

We note that, when computing the fit parameters, we perform a Chi-Squared analysis, taking the spectral uncertainties to be statistical.
For each energy bin ($E_i$), the fit residual is calculated:
\begin{equation}
\chi_{i}=\frac{\Phi_i-f\left(E_i\right)}{\sigma_i}\,,
\end{equation}
where ${\sigma_i}$ denotes the statistical measurement uncertainty at 68$\%$ confidence level. The value of $\chi^2$ can now easily be obtained by summing the residuals over all energy bins and squaring.

Assuming that $P_{\gamma \rightarrow \gamma}$ accurately predicts the Universe's opacity to VHE $\gamma$ rays, we would expect the residuals in the optically thick regime to follow a Gaussian distribution with a mean of zero. To test this hypothesis, we employ the $t$-test, where we calculate the test statistic, $t$, as follows:
\be
t=\frac{\bar{\chi}}{\sqrt{\sigma_\chi / N_\chi}}\,.
\ee
where  $\bar{\chi}$ denotes the mean and $\sigma_\chi$  the variance of the residual distribution, which comprises a total of $N_\chi$ data points. The $p_t$ test statistic we use to quantify the fit of our model to the observed spectra can be obtained using a standard t-test procedure.

\bibliographystyle{JHEP}
\bibliography{axionbib}

\providecommand{\href}[2]{#2}\begingroup\raggedright\begin{thebibliography}{10}

\bibitem{Preskill:1982cy}
J.~Preskill, M.~B. Wise, and F.~Wilczek, {\it {Cosmology of the Invisible
  Axion}},  {\em Phys. Lett. B} {\bf 120} (1983) 127--132.

\bibitem{Abbott:1982af}
L.~F. Abbott and P.~Sikivie, {\it {A Cosmological Bound on the Invisible
  Axion}},  {\em Phys. Lett. B} {\bf 120} (1983) 133--136.

\bibitem{Dine:1982ah}
M.~Dine and W.~Fischler, {\it {The Not So Harmless Axion}},  {\em Phys. Lett.
  B} {\bf 120} (1983) 137--141.

\bibitem{Marsh:2015xka}
D.~J.~E. Marsh, {\it {Axion Cosmology}},  {\em Phys. Rept.} {\bf 643} (2016)
  1--79, [\href{http://arxiv.org/abs/1510.07633}{{\tt arXiv:1510.07633}}].

\bibitem{Chadha-Day:2021szb}
F.~Chadha-Day, J.~Ellis, and D.~J.~E. Marsh, {\it {Axion dark matter: What is
  it and why now?}},  {\em Sci. Adv.} {\bf 8} (2022), no.~8 abj3618,
  [\href{http://arxiv.org/abs/2105.01406}{{\tt arXiv:2105.01406}}].

\bibitem{Peccei:1977hh}
R.~D. Peccei and H.~R. Quinn, {\it {CP Conservation in the Presence of
  Instantons}},  {\em Phys. Rev. Lett.} {\bf 38} (1977) 1440--1443.

\bibitem{Wilczek:1977pj}
F.~Wilczek, {\it {Problem of Strong $P$ and $T$ Invariance in the Presence of
  Instantons}},  {\em Phys. Rev. Lett.} {\bf 40} (1978) 279--282.

\bibitem{Voloshin:2006pz}
M.~B. Voloshin, {\it {Channel coupling in e+ e- annihilation into heavy meson
  pairs at the D* anti-D* threshold}},
  \href{http://arxiv.org/abs/hep-ph/0602233}{{\tt hep-ph/0602233}}.

\bibitem{Svrcek:2006yi}
P.~Svrcek and E.~Witten, {\it {Axions In String Theory}},  {\em JHEP} {\bf 06}
  (2006) 051, [\href{http://arxiv.org/abs/hep-th/0605206}{{\tt
  hep-th/0605206}}].

\bibitem{Cicoli:2012sz}
M.~Cicoli, M.~Goodsell, and A.~Ringwald, {\it {The type IIB string axiverse and
  its low-energy phenomenology}},  {\em JHEP} {\bf 10} (2012) 146,
  [\href{http://arxiv.org/abs/1206.0819}{{\tt arXiv:1206.0819}}].

\bibitem{Broeckel:2021dpz}
I.~Broeckel, M.~Cicoli, A.~Maharana, K.~Singh, and K.~Sinha, {\it {Moduli
  stabilisation and the statistics of axion physics in the landscape}},  {\em
  JHEP} {\bf 08} (2021) 059, [\href{http://arxiv.org/abs/2105.02889}{{\tt
  arXiv:2105.02889}}]. [Addendum: JHEP 01, 191 (2022)].

\bibitem{Demirtas:2021gsq}
M.~Demirtas, N.~Gendler, C.~Long, L.~McAllister, and J.~Moritz, {\it {PQ
  Axiverse}},  \href{http://arxiv.org/abs/2112.04503}{{\tt arXiv:2112.04503}}.

\bibitem{Gendler:2023kjt}
N.~Gendler, D.~J.~E. Marsh, L.~McAllister, and J.~Moritz, {\it {Glimmers from
  the Axiverse}},  \href{http://arxiv.org/abs/2309.13145}{{\tt
  arXiv:2309.13145}}.

\bibitem{Stott:2018opm}
M.~J. Stott and D.~J.~E. Marsh, {\it {Black hole spin constraints on the mass
  spectrum and number of axionlike fields}},  {\em Phys. Rev. D} {\bf 98}
  (2018), no.~8 083006, [\href{http://arxiv.org/abs/1805.02016}{{\tt
  arXiv:1805.02016}}].

\bibitem{Mehta:2021pwf}
V.~M. Mehta, M.~Demirtas, C.~Long, D.~J.~E. Marsh, L.~McAllister, and M.~J.
  Stott, {\it {Superradiance in string theory}},  {\em JCAP} {\bf 07} (2021)
  033, [\href{http://arxiv.org/abs/2103.06812}{{\tt arXiv:2103.06812}}].

\bibitem{Stott:2017hvl}
M.~J. Stott, D.~J.~E. Marsh, C.~Pongkitivanichkul, L.~C. Price, and B.~S.
  Acharya, {\it {Spectrum of the axion dark sector}},  {\em Phys. Rev. D} {\bf
  96} (2017), no.~8 083510, [\href{http://arxiv.org/abs/1706.03236}{{\tt
  arXiv:1706.03236}}].

\bibitem{Reig:2021ipa}
M.~Reig, {\it {The stochastic axiverse}},  {\em JHEP} {\bf 09} (2021) 207,
  [\href{http://arxiv.org/abs/2104.09923}{{\tt arXiv:2104.09923}}].

\bibitem{Kim:2004rp}
J.~E. Kim, H.~P. Nilles, and M.~Peloso, {\it {Completing natural inflation}},
  {\em JCAP} {\bf 01} (2005) 005,
  [\href{http://arxiv.org/abs/hep-ph/0409138}{{\tt hep-ph/0409138}}].

\bibitem{Dimopoulos:2005ac}
S.~Dimopoulos, S.~Kachru, J.~McGreevy, and J.~G. Wacker, {\it {N-flation}},
  {\em JCAP} {\bf 08} (2008) 003,
  [\href{http://arxiv.org/abs/hep-th/0507205}{{\tt hep-th/0507205}}].

\bibitem{Agrawal:2017cmd}
P.~Agrawal, J.~Fan, M.~Reece, and L.-T. Wang, {\it {Experimental Targets for
  Photon Couplings of the QCD Axion}},  {\em JHEP} {\bf 02} (2018) 006,
  [\href{http://arxiv.org/abs/1709.06085}{{\tt arXiv:1709.06085}}].

\bibitem{Kitajima:2014xla}
N.~Kitajima and F.~Takahashi, {\it {Resonant conversions of QCD axions into
  hidden axions and suppressed isocurvature perturbations}},  {\em JCAP} {\bf
  01} (2015) 032, [\href{http://arxiv.org/abs/1411.2011}{{\tt
  arXiv:1411.2011}}].

\bibitem{Obata:2021nql}
I.~Obata, {\it {Implications of the cosmic birefringence measurement for the
  axion dark matter search}},  {\em JCAP} {\bf 09} (2022) 062,
  [\href{http://arxiv.org/abs/2108.02150}{{\tt arXiv:2108.02150}}].

\bibitem{Higaki:2014qua}
T.~Higaki, N.~Kitajima, and F.~Takahashi, {\it {Hidden axion dark matter
  decaying through mixing with QCD axion and the 3.5 keV X-ray line}},  {\em
  JCAP} {\bf 12} (2014) 004, [\href{http://arxiv.org/abs/1408.3936}{{\tt
  arXiv:1408.3936}}].

\bibitem{Glennon:2023jsp}
N.~Glennon, N.~Musoke, and C.~Prescod-Weinstein, {\it {Simulations of
  multifield ultralight axionlike dark matter}},  {\em Phys. Rev. D} {\bf 107}
  (2023), no.~6 063520, [\href{http://arxiv.org/abs/2302.04302}{{\tt
  arXiv:2302.04302}}].

\bibitem{Gavela:2023tzu}
B.~Gavela, P.~Qu\'\i{}lez, and M.~Ramos, {\it {Multiple QCD axion}},
  \href{http://arxiv.org/abs/2305.15465}{{\tt arXiv:2305.15465}}.

\bibitem{Chadha-Day:2021uyt}
F.~Chadha-Day, {\it {Axion-like particle oscillations}},  {\em JCAP} {\bf 01}
  (2022), no.~01 013, [\href{http://arxiv.org/abs/2107.12813}{{\tt
  arXiv:2107.12813}}].

\bibitem{Chen:2021hfq}
Z.~Chen, A.~Kobakhidze, C.~A.~J. O'Hare, Z.~S.~C. Picker, and G.~Pierobon, {\it
  {Phenomenology of the companion-axion model: photon couplings}},  {\em Eur.
  Phys. J. C} {\bf 82} (2022), no.~10 940,
  [\href{http://arxiv.org/abs/2109.12920}{{\tt arXiv:2109.12920}}].

\bibitem{Cyncynates:2021xzw}
D.~Cyncynates, T.~Giurgica-Tiron, O.~Simon, and J.~O. Thompson, {\it {Resonant
  nonlinear pairs in the axiverse and their late-time direct and astrophysical
  signatures}},  {\em Phys. Rev. D} {\bf 105} (2022), no.~5 055005,
  [\href{http://arxiv.org/abs/2109.09755}{{\tt arXiv:2109.09755}}].

\bibitem{Gasparotto:2023psh}
S.~Gasparotto and E.~I. Sfakianakis, {\it {Cosmic Birefringence from the
  Axiverse}},  \href{http://arxiv.org/abs/2306.16355}{{\tt arXiv:2306.16355}}.

\bibitem{Cyncynates:2022wlq}
D.~Cyncynates, O.~Simon, J.~O. Thompson, and Z.~J. Weiner, {\it
  {Nonperturbative structure in coupled axion sectors and implications for
  direct detection}},  {\em Phys. Rev. D} {\bf 106} (2022), no.~8 083503,
  [\href{http://arxiv.org/abs/2208.05501}{{\tt arXiv:2208.05501}}].

\bibitem{Cyncynates:2023esj}
D.~Cyncynates and J.~O. Thompson, {\it {Heavy QCD axion dark matter from
  avoided level crossing}},  {\em Phys. Rev. D} {\bf 108} (2023), no.~9
  L091703, [\href{http://arxiv.org/abs/2306.04678}{{\tt arXiv:2306.04678}}].

\bibitem{Dienes:1999gw}
K.~R. Dienes, E.~Dudas, and T.~Gherghetta, {\it {Invisible axions and large
  radius compactifications}},  {\em Phys. Rev. D} {\bf 62} (2000) 105023,
  [\href{http://arxiv.org/abs/hep-ph/9912455}{{\tt hep-ph/9912455}}].

\bibitem{Ayala:2014pea}
A.~Ayala, I.~Dom\'\i{}nguez, M.~Giannotti, A.~Mirizzi, and O.~Straniero, {\it
  {Revisiting the bound on axion-photon coupling from Globular Clusters}},
  {\em Phys. Rev. Lett.} {\bf 113} (2014), no.~19 191302,
  [\href{http://arxiv.org/abs/1406.6053}{{\tt arXiv:1406.6053}}].

\bibitem{Raffelt:1985nj}
G.~G. Raffelt, {\it {Axion Constraints From White Dwarf Cooling Times}},  {\em
  Phys. Lett. B} {\bf 166} (1986) 402--406.

\bibitem{Blinnikov:1994eoa}
S.~I. Blinnikov and N.~V. Dunina-Barkovskaya, {\it {The cooling of hot white
  dwarfs: a theory with non-standard weak interactions, and a comparison with
  observations}},  {\em Mon. Not. Roy. Astron. Soc.} {\bf 266} (1994) 289--304.

\bibitem{Halverson:2018cio}
J.~Halverson and F.~Ruehle, {\it {Computational Complexity of Vacua and
  Near-Vacua in Field and String Theory}},  {\em Phys. Rev. D} {\bf 99} (2019),
  no.~4 046015, [\href{http://arxiv.org/abs/1809.08279}{{\tt
  arXiv:1809.08279}}].

\bibitem{Halverson:2019cmy}
J.~Halverson, C.~Long, B.~Nelson, and G.~Salinas, {\it {Towards string theory
  expectations for photon couplings to axionlike particles}},  {\em Phys. Rev.
  D} {\bf 100} (2019), no.~10 106010,
  [\href{http://arxiv.org/abs/1909.05257}{{\tt arXiv:1909.05257}}].

\bibitem{Hall:1999sn}
L.~J. Hall, H.~Murayama, and N.~Weiner, {\it {Neutrino mass anarchy}},  {\em
  Phys. Rev. Lett.} {\bf 84} (2000) 2572--2575,
  [\href{http://arxiv.org/abs/hep-ph/9911341}{{\tt hep-ph/9911341}}].

\bibitem{Haba:2000be}
N.~Haba and H.~Murayama, {\it {Anarchy and hierarchy}},  {\em Phys. Rev. D}
  {\bf 63} (2001) 053010, [\href{http://arxiv.org/abs/hep-ph/0009174}{{\tt
  hep-ph/0009174}}].

\bibitem{deGouvea:2003xe}
A.~de~Gouvea and H.~Murayama, {\it {Statistical test of anarchy}},  {\em Phys.
  Lett. B} {\bf 573} (2003) 94--100,
  [\href{http://arxiv.org/abs/hep-ph/0301050}{{\tt hep-ph/0301050}}].

\bibitem{deGouvea:2012ac}
A.~de~Gouvea and H.~Murayama, {\it {Neutrino Mixing Anarchy: Alive and
  Kicking}},  {\em Phys. Lett. B} {\bf 747} (2015) 479--483,
  [\href{http://arxiv.org/abs/1204.1249}{{\tt arXiv:1204.1249}}].

\bibitem{Espinosa:2003qz}
J.~R. Espinosa, {\it {Anarchy in the neutrino sector?}},
  \href{http://arxiv.org/abs/hep-ph/0306019}{{\tt hep-ph/0306019}}.

\bibitem{Heeck:2012fw}
J.~Heeck, {\it {Seesaw parametrization for n right-handed neutrinos}},  {\em
  Phys. Rev. D} {\bf 86} (2012) 093023,
  [\href{http://arxiv.org/abs/1207.5521}{{\tt arXiv:1207.5521}}].

\bibitem{Bai:2012zn}
Y.~Bai and G.~Torroba, {\it {Large N (=3) Neutrinos and Random Matrix Theory}},
   {\em JHEP} {\bf 12} (2012) 026, [\href{http://arxiv.org/abs/1210.2394}{{\tt
  arXiv:1210.2394}}].

\bibitem{Lu:2014cla}
X.~Lu and H.~Murayama, {\it {Neutrino Mass Anarchy and the Universe}},  {\em
  JHEP} {\bf 08} (2014) 101, [\href{http://arxiv.org/abs/1405.0547}{{\tt
  arXiv:1405.0547}}].

\bibitem{Fortin:2016zyf}
J.-F. Fortin, N.~Giasson, and L.~Marleau, {\it {Probability density function
  for neutrino masses and mixings}},  {\em Phys. Rev. D} {\bf 94} (2016),
  no.~11 115004, [\href{http://arxiv.org/abs/1609.08581}{{\tt
  arXiv:1609.08581}}].

\bibitem{Fortin:2017iiw}
J.-F. Fortin, N.~Giasson, and L.~Marleau, {\it {Anarchy and Neutrino Physics}},
   {\em JHEP} {\bf 04} (2017) 131, [\href{http://arxiv.org/abs/1702.07273}{{\tt
  arXiv:1702.07273}}].

\bibitem{Fortin:2018etr}
J.-F. Fortin, N.~Giasson, and L.~Marleau, {\it {Probability Density Functions
  for CP-Violating Rephasing Invariants}},  {\em Nucl. Phys. B} {\bf 930}
  (2018) 384--398, [\href{http://arxiv.org/abs/1801.10165}{{\tt
  arXiv:1801.10165}}].

\bibitem{Fortin:2020oud}
J.-F. Fortin, N.~Giasson, L.~Marleau, and J.~Pelletier-Dumont, {\it {Mellin
  transform approach to rephasing invariants}},  {\em Phys. Rev. D} {\bf 102}
  (2020), no.~3 036001, [\href{http://arxiv.org/abs/2004.14284}{{\tt
  arXiv:2004.14284}}].

\bibitem{Raffelt:1985nk}
G.~G. Raffelt, {\it {ASTROPHYSICAL AXION BOUNDS DIMINISHED BY SCREENING
  EFFECTS}},  {\em Phys. Rev. D} {\bf 33} (1986) 897.

\bibitem{Krauss:1984gm}
L.~M. Krauss, J.~E. Moody, and F.~Wilczek, {\it {A STELLAR ENERGY LOSS
  MECHANISM INVOLVING AXIONS}},  {\em Phys. Lett. B} {\bf 144} (1984) 391--394.

\bibitem{Dimopoulos:1985tm}
S.~Dimopoulos, G.~D. Starkman, and B.~W. Lynn, {\it {Atomic Enhancements in the
  Detection of Weakly Interacting Particles}},  {\em Phys. Lett. B} {\bf 168}
  (1986) 145--150.

\bibitem{Dimopoulos:1986kc}
S.~Dimopoulos, J.~A. Frieman, B.~W. Lynn, and G.~D. Starkman, {\it
  {Axiorecombination: A New Mechanism for Stellar Axion Production}},  {\em
  Phys. Lett. B} {\bf 179} (1986) 223--227.

\bibitem{Pospelov:2008jk}
M.~Pospelov, A.~Ritz, and M.~B. Voloshin, {\it {Bosonic super-WIMPs as
  keV-scale dark matter}},  {\em Phys. Rev. D} {\bf 78} (2008) 115012,
  [\href{http://arxiv.org/abs/0807.3279}{{\tt arXiv:0807.3279}}].

\bibitem{Mikaelian:1978jg}
K.~O. Mikaelian, {\it {Astrophysical Implications of New Light Higgs Bosons}},
  {\em Phys. Rev. D} {\bf 18} (1978) 3605.

\bibitem{Fukugita:1982ep}
M.~Fukugita, S.~Watamura, and M.~Yoshimura, {\it {Light Pseudoscalar Particle
  and Stellar Energy Loss}},  {\em Phys. Rev. Lett.} {\bf 48} (1982) 1522.

\bibitem{Fukugita:1982gn}
M.~Fukugita, S.~Watamura, and M.~Yoshimura, {\it {Astrophysical Constraints on
  a New Light Axion and Other Weakly Interacting Particles}},  {\em Phys. Rev.
  D} {\bf 26} (1982) 1840.

\bibitem{Redondo:2013wwa}
J.~Redondo, {\it {Solar axion flux from the axion-electron coupling}},  {\em
  JCAP} {\bf 12} (2013) 008, [\href{http://arxiv.org/abs/1310.0823}{{\tt
  arXiv:1310.0823}}].

\bibitem{Irastorza:2011gs}
I.~G. Irastorza et~al., {\it {Towards a new generation axion helioscope}},
  {\em JCAP} {\bf 06} (2011) 013, [\href{http://arxiv.org/abs/1103.5334}{{\tt
  arXiv:1103.5334}}].

\bibitem{Hoof:2021mld}
S.~Hoof, J.~Jaeckel, and L.~J. Thormaehlen, {\it {Quantifying uncertainties in
  the solar axion flux and their impact on determining axion model
  parameters}},  {\em JCAP} {\bf 09} (2021) 006,
  [\href{http://arxiv.org/abs/2101.08789}{{\tt arXiv:2101.08789}}].

\bibitem{Barth:2013sma}
K.~Barth et~al., {\it {CAST constraints on the axion-electron coupling}},  {\em
  JCAP} {\bf 05} (2013) 010, [\href{http://arxiv.org/abs/1302.6283}{{\tt
  arXiv:1302.6283}}].

\bibitem{Gondolo:2008dd}
P.~Gondolo and G.~G. Raffelt, {\it {Solar neutrino limit on axions and keV-mass
  bosons}},  {\em Phys. Rev. D} {\bf 79} (2009) 107301,
  [\href{http://arxiv.org/abs/0807.2926}{{\tt arXiv:0807.2926}}].

\bibitem{Dessert:2019sgw}
C.~Dessert, A.~J. Long, and B.~R. Safdi, {\it {X-ray Signatures of Axion
  Conversion in Magnetic White Dwarf Stars}},  {\em Phys. Rev. Lett.} {\bf 123}
  (2019), no.~6 061104, [\href{http://arxiv.org/abs/1903.05088}{{\tt
  arXiv:1903.05088}}].

\bibitem{doi:10.1093/pasj/65.4.73}
A.~Harayama, Y.~Terada, M.~Ishida, T.~Hayashi, A.~Bamba, and M.~S. Tashiro,
  {\it Search for non-thermal emissions from an isolated magnetic white dwarf,
  euve j0317 855, with suzaku},  {\em Publications of the Astronomical Society
  of Japan} {\bf 65} (2013), no.~4 73.

\bibitem{Kulebi:2010pd}
B.~Kulebi, S.~Jordan, E.~Nelan, U.~Bastian, and M.~Altmann, {\it {Constraints
  on the origin of the massive, hot, and rapidly rotating magnetic white dwarf
  RE J 0317-853 from an HST parallax measurement}},  {\em Astron. Astrophys.}
  {\bf 524} (2010) A36, [\href{http://arxiv.org/abs/1007.4978}{{\tt
  arXiv:1007.4978}}].

\bibitem{2014JETPL}
G.~I. {Rubtsov} and S.~V. {Troitsky}, {\it {Breaks in gamma-ray spectra of
  distant blazars and transparency of the Universe}},  {\em Soviet Journal of
  Experimental and Theoretical Physics Letters} {\bf 100} (Nov., 2014)
  355--359, [\href{http://arxiv.org/abs/1406.0239}{{\tt arXiv:1406.0239}}].

\bibitem{Kohri_2017}
K.~Kohri and H.~Kodama, {\it Axion-like particles and recent observations of
  the cosmic infrared background radiation},  {\em Physical Review D} {\bf 96}
  (sep, 2017).

\bibitem{Biteau_2015}
J.~Biteau and D.~A. Williams, {\it {THE} {EXTRAGALACTIC} {BACKGROUND} {LIGHT},
  {THE} {HUBBLE} {CONSTANT}, {AND} {ANOMALIES}: {CONCLUSIONS} {FROM} 20 {YEARS}
  {OF} {TeV} {GAMMA}-{RAY} {OBSERVATIONS}},  {\em The Astrophysical Journal}
  {\bf 812} (oct, 2015) 60.

\bibitem{Dominguez:2015ama}
A.~Dom\'\i{}nguez and M.~Ajello, {\it {Spectral analysis of Fermi-LAT blazars
  above 50 GeV}},  {\em Astrophys. J. Lett.} {\bf 813} (2015), no.~2 L34,
  [\href{http://arxiv.org/abs/1510.07913}{{\tt arXiv:1510.07913}}].

\bibitem{DeAngelis:2011id}
A.~De~Angelis, G.~Galanti, and M.~Roncadelli, {\it {Relevance of axion-like
  particles for very-high-energy astrophysics}},  {\em Phys. Rev. D} {\bf 84}
  (2011) 105030, [\href{http://arxiv.org/abs/1106.1132}{{\tt
  arXiv:1106.1132}}]. [Erratum: Phys.Rev.D 87, 109903 (2013)].

\bibitem{Simet_2008}
M.~Simet, D.~Hooper, and P.~D. Serpico, {\it Milky way as a kiloparsec-scale
  axionscope},  {\em Physical Review D} {\bf 77} (mar, 2008).

\bibitem{Sanchez-Conde:2009exi}
M.~A. Sanchez-Conde, D.~Paneque, E.~Bloom, F.~Prada, and A.~Dominguez, {\it
  {Hints of the existence of Axion-Like-Particles from the gamma-ray spectra of
  cosmological sources}},  {\em Phys. Rev. D} {\bf 79} (2009) 123511,
  [\href{http://arxiv.org/abs/0905.3270}{{\tt arXiv:0905.3270}}].

\bibitem{Meyer_2013}
M.~Meyer, D.~Horns, and M.~Raue, {\it First lower limits on the
  photon-axion-like particle coupling from very high energy gamma-ray
  observations},  {\em Physical Review D} {\bf 87} (feb, 2013).

\bibitem{Masaki_2017}
E.~Masaki, A.~Aoki, and J.~Soda, {\it Photon-axion conversion, magnetic field
  configuration, and polarization of photons},  {\em Physical Review D} {\bf
  96} (aug, 2017).

\bibitem{Franceschini_2008}
A.~Franceschini, G.~Rodighiero, and M.~Vaccari, {\it Extragalactic
  optical-infrared background radiation, its time evolution and the cosmic
  photon-photon opacity},  {\em Astronomy {\&} Astrophysics} {\bf 487} (jun,
  2008) 837--852.

\bibitem{Jansson_2012}
R.~Jansson and G.~R. Farrar, {\it A {NEW} {MODEL} {OF} {THE} {GALACTIC}
  {MAGNETIC} {FIELD}},  {\em The Astrophysical Journal} {\bf 757} (aug, 2012)
  14.

\bibitem{ADMX:2018ogs}
{\bf ADMX} Collaboration, C.~Boutan et~al., {\it {Piezoelectrically Tuned
  Multimode Cavity Search for Axion Dark Matter}},  {\em Phys. Rev. Lett.} {\bf
  121} (2018), no.~26 261302, [\href{http://arxiv.org/abs/1901.00920}{{\tt
  arXiv:1901.00920}}].

\bibitem{Meyer:2013mma}
M.~Meyer, {\em {The Opacity of the Universe for High and Very High Energy
  $\gamma$-Rays}}.
\newblock PhD thesis, Hamburg U., 2013.

\end{thebibliography}\endgroup
% \printbibliography

\end{document}